%% file: ijcai25.tex
\documentclass{article}
\usepackage{ijcai25}

\usepackage{times}
\usepackage{soul}
\usepackage{url}
\usepackage[utf8]{inputenc}
\usepackage[small]{caption}
\usepackage{graphicx}
\usepackage{amsmath}
\usepackage{amsthm}
\usepackage{booktabs}
\usepackage{algorithm}
\usepackage{algorithmic}
\usepackage[switch]{lineno}

\title{Can Humans Teach Machines to Code?}

\author {
    Céline Hocquette\textsuperscript{\rm 1}\and
    Johannes Langer\textsuperscript{\rm 2}\and
    Andrew Cropper\textsuperscript{\rm 3}\and
    Ute Schmid\textsuperscript{\rm 2}
\affiliations
    \textsuperscript{\rm 1}University of Southampton\\
    \textsuperscript{\rm 2}University of Bamberg\\
    \textsuperscript{\rm 3}University of Oxford\\
\emails
c.m.e.j.hocquette@soton.ac.uk
}

\input{preamble}

\input{data}
\usepackage[toc,page]{appendix}

\begin{document}

\maketitle

\input{00-abstract}

\input{01-intro}
\input{02-related}

\input{04-experiments}
\input{06-conclusions}

\bibliographystyle{named}
\bibliography{ijcai25}

\begin{appendices}
 
This appendix contains supplementary material for the paper \emph{Can humans teach machines to code?}.
The outline is as follows:

\begin{itemize}
\item Section \ref{sec:prompts} shows our prompts for \textsc{DeepSeek-Coder-V2}.
\item Section \ref{sec:interface} shows our experimental interface.
% \item Section \ref{sec:targets} provides an example program for each of the tasks considered.
\item Section \ref{sec:expertexs} shows the examples provided by the expert.
\item Section \ref{sec:appendixresults} shows the detail of our results for each of the tasks.
\end{itemize}

\input{prompts}
\input{A-interface}
% \input{A-targets}
\input{B-expertexs}

\input{C-results}

\end{appendices}


\end{document}

%% file: preamble.tex
\usepackage[utf8]{inputenc}
\usepackage{xcolor}
\usepackage{inconsolata}
\usepackage{tikz}
\usetikzlibrary{patterns}
\usepackage{pgfplots}
\usepackage{microtype}
\usepackage{amssymb}
\usepackage{algorithm}
\usepackage{subfig}

\usepackage{multirow}
\usepackage{amsmath}
\usepackage{amsthm}
\usepackage{rotating}
\usepackage{footnote}
\makesavenoteenv{tabular}
\makesavenoteenv{table}

\usepackage{booktabs}

\theoremstyle{definition}

\definecolor{mygreen}{rgb}{0.3,0.6,0.4}

\newcommand{\popper}{\textsc{Popper}}
\newcommand{\ale}{\textsc{aleph}}
\newcommand{\metagol}{\textsc{metagol}}
\usepackage{listings}
\usepackage{hyperref}

\usepackage{pgfplots}

\usepgfplotslibrary{colorbrewer}
\pgfplotsset{compat = 1.15, cycle list/Set1-8} 
\usetikzlibrary{pgfplots.statistics, pgfplots.colorbrewer} 
\usepackage{pgfplotstable}
\lstnewenvironment{myalgorithm}[1][] %defines the algorithm listing environment
{
    \lstset{ %this is the stype
        basicstyle=\ttfamily\small,
        columns=flexible,
        showspaces=false,               % show spaces adding particular underscores
        showstringspaces=false,         % underline spaces within strings
        mathescape=true,
        numbers=left,
        escapeinside={*}{*},
        columns=flexible,
        escapechar=@,
        numbersep=3pt,        % default 10pt
        keywordstyle=\bfseries,
        keywords={,and, return, not, def, in,elif,if, else, for, foreach, while,}
        numbers=left,
        xleftmargin=8pt,
        #1 % this is to add specific settings to an usage of this environment (for instnce, the caption and referable label)
    }
}
{}

\usepackage{pgfkeys}
    \newenvironment{customlegend}[1][]{%
        \begingroup
        \csname pgfplots@init@cleared@structures\endcsname
        \pgfplotsset{#1}%
    }{%
        \csname pgfplots@createlegend\endcsname
        \endgroup
    }%
    \def\addlegendimage{\scriptsize\csname pgfplots@addlegendimage\endcsname}
\usepackage{comment}

%% file: 00-abstract.tex
\begin{abstract}
The goal of inductive program synthesis is for a machine to automatically generate a program from user-supplied examples. 
A key underlying assumption is that humans can provide sufficient examples to teach a concept to a machine.
To evaluate the validity of this assumption, we conduct a study where human participants provide examples for six programming concepts, such as finding the maximum element of a list.
We evaluate the generalisation performance of five program synthesis systems trained on input-output examples (i) from non-expert humans, (ii) from a human expert, and (iii) randomly sampled.
% \jl{We evaluate the generalisation performance of 5 different approaches to program synthesis, including three Inductive Logic Programming systems, one Bayesian inductive learning system, and one code-trained Large Language Model. Each system was trained on input-output-examples from (i) non-expert humans, (ii) a human expert, and (iii) randomly sampled.}
% \ac{it is not a random source, it is randomly generated by us}.
Our results suggest that non-experts typically do not provide sufficient examples for a program synthesis system to learn an accurate program.
% Additionally, non-experts require more examples than both the expert and those randomly sampled to achieve comparable accuracies.
% \ch{this last sentence is a bit akwardly phrased}
% \ac{I agree, cut?}
\end{abstract}

%% file: 01-intro.tex
\section{Introduction}

Synthesising a computer program from an incomplete specification is known as \emph{(inductive) program synthesis}\footnote{We acknowledge there are other forms of program synthesis, such as deductive program synthesis \cite{manna1971toward}. We focus on \emph{inductive} program synthesis.} and is a grand challenge in artificial intelligence \cite{flener2008introduction,cacm:ip}.
In \emph{programming by example} (PBE), the specification consists of input-output examples demonstrating the desired program's behaviour. 
Examples are typically provided by humans with little programming experience \cite{lieberman:pbe}.
% Specifying a programming problem by a set of input-output examples in general is more intuitive to humans than to provide complete natural language or formal specifications \cite{schmid2011inductive,cacm:ip}.

Program synthesis can be viewed as a machine learning problem \cite{tm:book}.
However, while standard machine learning approaches learn an attribute-value model \cite{luc:book}, program synthesis approaches learn a computer program, such as a LISP \cite{summers:lisp}, Prolog \cite{mis}, 
or Haskell \cite{magichaskeller} program.

% \begin{figure}
%     \centering
%     \includegraphics[width=\linewidth]{introduction_illustration@4x.png}
%     \caption{Program synthesis in practice. A human provides a set of input-output example of a desired program's behaviour to a program synthesis system, which returns a program.
%     \ac{I would write the rules in natural language or in Python. It does not matter that the rules are logic programs. Most readers will not know logic programs.}
%     \ch{write the program in several languages, Python, Prolog and maybe other or natural language}
%     \ac{Also, if we are short for space, I would cut this illustration. I think we can describe program synthesis clearly enough without it.}
%     }
%     \label{fig:illustration}
% \end{figure}

% REVISED
As with all machine learning paradigms, the performance of a program synthesis system heavily depends on the quality of its training examples \cite{winston1970learning,flach2012machine}.
% To achieve good performance, training examples must be representative of the concept being taught.
While most standard machine learning approaches rely on large amounts of training data to ensure good performance, obtaining thousands of examples is often impractical in program synthesis applications where examples are user-provided. 
For instance, FlashFill \cite{flashfill}, a PBE system in Microsoft Excel, induces string transformation programs from user-provided examples, such as transforming \emph{``Alan Mathison Turing''} to \emph{``A.M. Turing''}.
In such scenarios, expecting human users to provide thousands of examples is unrealistic.

A key assumption in user-driven program synthesis is that humans provide sufficient examples for a synthesis system to learn the desired concept \cite{lieberman1986example,lau2009programming}.
However, to the best of our knowledge, this assumption lacks both empirical evidence and theoretical support.
In other words, it remains an open question whether humans provide sufficient examples for program synthesis systems to generalise effectively.

To address this gap, our goal is to answer the question \emph{``Do humans provide sufficient examples to teach a programming concept to a program synthesis system?''}. 
By `sufficient', we mean that a synthesis system learns a program that generalises with high accuracy to unseen examples.
This question differs from prior studies, which focus on whether humans teach optimally in terms of number of examples \cite{good_teaching}.
Instead, we investigate whether human-provided examples are sufficient for a synthesis system to learn the desired concepts. 
Additionally, we focus on teaching program synthesis systems rather than traditional machine learning classifiers.

% \footnote{
% See \citeauthor{rulehacker} [\citeyear{rulehacker}] (Chapter 3) for more background on list concepts.
% }.
% For instance, suppose you want to teach a machine the concept \emph{``return the last element of a list''}.
% Then a human teacher might provide examples such as 
% \emph{[a,l,i,c,e] $\mapsto$ e}, 
% \emph{[j,i,m] $\mapsto$ m}, 
% \emph{[s,a,l,l,y] $\mapsto$ y}.
% Given these examples, a program synthesis system should generate a program that correctly generalises the training examples and, crucially, generalises to unseen examples. 

To answer this question, we conduct an empirical study where we ask human participants to provide examples to teach programming concepts\footnote{No ethics statement was required by the home university for this empirical study.
% required for submission
}. 
Unlike studies that evaluate the effects of teaching guidance \cite{good_teaching}, we do not give participants teaching instructions. We evaluate whether humans naturally provide sufficient examples, rather than determining whether a setting exists in which they could do so.
This choice reflects real-world scenarios where users interact with synthesis systems in unconstrained environments without explicit instructions on how to provide examples.
We collect examples from participants with diverse backgrounds: (i) non-computer scientists, and (ii) computer scientists without expertise in machine learning or program synthesis. Additionally, we use examples from one expert in program synthesis as a baseline.
% We hypothesise the expert knows the best examples to provide.
We evaluate multiple synthesis systems on these examples, including inductive logic programming (ILP) systems, one Bayesian inductive learning system, and one code-trained large language model (LLM).
For each participant, we train each synthesis system on their examples and evaluate the predictive performance of the learned program. 

Our study focuses on teaching recursive list manipulation concepts.
Lists are a core data structure in computer science and have many applications, such as in computational biology, where proteins, genes, and DNA are typically represented as strings \cite{luc:book}.
%Moreover, list functions are a standard benchmark for program synthesis \cite{summers:lisp,mis,hofmann2009unifying,flashfill,metabias,dreamcoder,metaopt} \jl{\cite{dagstuhl2024}} and human learning \cite{RULE2020900,ruleefficient}.
Moreover, list functions are a standard benchmark for human learning \cite{RULE2020900,ruleefficient} and program synthesis \cite{summers:lisp,mis,hofmann2009unifying,flashfill,metabias,dreamcoder,metaopt,dagstuhl2024}, including modern LLM-based code generators \cite{codex}.
% \ac{needs a better ending}

\subsubsection*{Novelty and Contributions}
Our main contribution is an empirical study that investigates whether humans can teach machines to code. 
As far as we are aware, our study is the first to evaluate whether human-provided examples are sufficient to train program synthesis systems.
Our results show two key insights.
\begin{enumerate}
\item A common assumption in program synthesis (and machine learning) is that human-provided examples are sufficient to teach a concept to a machine.
% Many program synthesis (and machine learning) researchers work under the assumption that humans are capable of providing sufficient examples to teach a concept to a machine.
Our empirical results challenge this assumption and show that, for most of the concepts evaluated, non-expert participants do not provide sufficient examples for any of the synthesis systems tested to learn an accurate program.
This negative result has potential implications for end-user applications of program synthesis, such as teaching home robot assistants.
% For instance, if using program synthesis to teach a home robot assistant to perform a task, our results suggest that a human could need to provide more examples than they would prefer to achieve the desired functionality.
\item Our results show that synthesis systems perform better when trained on randomly generated examples compared to those provided by non-expert humans.
Since many program synthesis (and machine learning) algorithms are optimised using synthetic datasets, this result suggests that synthesis systems might struggle when trained on human-provided examples, highlighting potential limitations for deploying synthesis systems in user-driven applications.
\end{enumerate}

%% file: 02-related.tex
\section{Related Work}
\label{sec:related}

\textbf{Program synthesis.}
% Neural approaches to program synthesis often require many examples \cite{nandopoo,}. 
LLMs have been widely used to generate programs from a natural language description \cite{codex,alphacode,coderl}, or from a combination of a description and a few examples \cite{polosukhin2018neural,austin2021program}. By contrast, we learn programs using only examples. Recent studies suggest that pretrained LLMs struggle to synthesise programs from just a few examples \cite{llmprogramsynthesis}. Although fine-tuning can improve their performance, LLMs still have difficulties with out-of-distribution problems.

\textbf{Machine teaching.}
Machine teaching \cite{machineteaching,telle2019teaching} is the problem of finding an optimal (usually minimal in size) training set that allows a learner to uniquely identify a target concept.
% The teaching dimension of a concept class is the minimum number of examples a teacher must provide for the learner to uniquely identify any target concept chosen from the class \cite{goldman1995complexity}. 
% They \ac{they = goldman or telle?} 
% \citeauthor{goldman1995complexity} [\citeyear{goldman1995complexity}] 
% show that the problem of finding an optimal teaching sequence for an arbitrary concept i.e. a minimal sequence of examples, 
% is NP-hard.
While most research in machine teaching is theoretical, we present empirical results involving humans. Instead of searching for an optimal training set, we examine whether humans naturally provide sufficient examples to teach concepts.
Machine teaching considers both humans and machines as teachers and learners \cite{zhu2018overview}, while we focus on humans teaching machines. 

\textbf{Human teaching humans.} 
% In human-to-human concept teaching, it is generally assumed that the teacher has full knowledge of a concept.
% Their goal is to choose data to facilitate learning.
%\ac{Does this sentence refer to 'In pedagogical reasoning'?}.
\citeauthor{shafto2014rational} [\citeyear{shafto2014rational}] show that human teachers select examples in a non-random manner, and prefer those that represent the mean and extent of the underlying distribution, which improves learning effectiveness.
% For instance, when teaching prototype concepts, teachers purposefully select examples that represent the mean and extent of the true distribution, rather than random examples.
% This purposeful selection improves the effectiveness of the learning process compared to random examples.
Our work considers machine learners instead of human learners.
% The learner in particular understand the difference between pedagogical and non-pedagogical situations, and use this knowledge to guide inference. 

% S. Goldman, M. Kearns, On the complexity of teaching, J. Comput. Syst. Sci. 50 (1) (February 1995) 20–31.

% investigate the acquisition of recursive programming skills in humans, and found that almost all students develop new programs by analogy to example recursive programs but that their success is determined by how well they understood why these examples work \cite{pirolli1985role}.
% \ac{ijcai game on game board ordering}

% \textbf{Machine Teaching.}
% \ac{jose's work}
% multi-task https://arxiv.org/abs/2009.06227

% \textbf{Algorithmic teaching.}
% The field of algorithmic teaching [3,29,15] studies the teaching problem: producing a set of labelled examples based on a known target concept. 
% The goal is to find efficient algorithms that can teach with as few examples as possible.

%\textbf{Human teaching humans.} 
%In the context of human-to-human teaching, empirical evidence suggests that teachers select examples which they consider to be most helpful to the information needs of a human learner \cite{shafto2014rational}.

\textbf{Human teaching machines.} 
Studies on how humans teach machines suggest that natural human teaching is often suboptimal in terms of the number of examples provided \cite{khan2011humans,good_teaching}. 
\citeauthor{khan2011humans} [\citeyear{khan2011humans}] empirically show that human teachers often start with extreme examples, rather than those near the decision boundary, which aligns with the curriculum learning principle.
While this study asks participants to select examples from a small predefined set, we ask participants to generate examples.
In contrast to previous works on sequential decision-making \cite{cakmak2012algorithmic} and binary classification tasks \cite{good_teaching}, our study focuses on program synthesis concepts involving recursion. 
Furthermore, while previous studies evaluate the impact of teaching guidance \cite{good_teaching,cakmak2012algorithmic}, we provide little teaching guidance to participants and evaluate whether humans naturally provide sufficient examples. 

%\ch{it is a experimental study of the teaching strategies employed by human teaching machines. It shows human teachers do not provide a minimal set of examples. Participants did not provide examples near the boundary (machine teaching strategy) but rather extreme examples, as predicted by curriculum learning. The target concept is to teach a 1D threshold to a robot.}
% \ac{The humans mostly pick examples from a set. They do not generate them themselves, except for in Experiment 4}
% \ch{which learning system did they use?}

\textbf{List functions.}
Lists are a simple yet rich domain.
As \citeauthor{rulehacker} \shortcite{rulehacker} explains, lists use numbers in multiple roles (symbols, ordinals, and cardinals) and their recursive structure supports the expression of complex concepts.  
% Recursion is crucial to generalise from a small number of examples to lists of arbitrary size \ac{citation?}. 
Lists naturally align with familiar psychological concepts and are commonly used in classic concept learning tasks.
However, recursion is cognitively challenging compared to natural categories for which humans can be considered experts \cite{kahney1983novice}. 
% , and are formally tractable
% \ac{what does formally tractable mean?}
% List problems are a standard inductive programming benchmark \cite{dagstuhl2024}. 
\citeauthor{rulehacker} [\citeyear{rulehacker}] uses list transformation problems to compare the performance of humans with program synthesis approaches.
However, their work does not address how effectively humans can teach a program synthesis system a concept.

%% file: 04-experiments.tex
\section{Empirical Study}
\label{sec:exp}

We conduct an empirical study to explore whether humans naturally provide sufficient examples for a program synthesis system to learn the desired concept.
Our primary question is:
\begin{description}
\item[Q1.] Do humans provide sufficient examples to teach a programming concept to a program synthesis system?
\end{description}
\noindent
% \ac{programming task or programming concept? let's only use one}
To answer \textbf{Q1}, we ask human participants to provide examples to teach programming concepts. We separately train various program synthesis systems on these examples, and evaluate the predictive accuracy of the learned programs on unseen test data. Unlike previous studies that evaluate the impact of teaching guidance on teaching performance \cite{good_teaching}, we do not give any instructions on the type of examples to provide. Instead, we evaluate whether humans, without guidance, naturally provide examples from which a program synthesis system generalises the desired concept.
% In this open-ended context, rather than 
% \ac{odd and confusing sentence. Break into two sentences}
% \ac{this "naturally supply" statement is quite important. I would stress this emphasis in the introduction}

We hypothesise that familiarity with the problem domain might help humans choose better examples. For instance, individuals with programming experience might have a deeper understanding of edge cases, such as the empty list for list manipulation concepts.
Therefore, our second question is:
\begin{description}
\item[Q2.] Does having a background in computer science improve a human's ability to teach a programming concept to a program synthesis system?
\end{description}
To answer \textbf{Q2}, we compare the performance of program synthesis systems trained on examples provided by participants with (i) no computer science experience (NCS), and (ii) computer science education but without specific expertise in program synthesis or machine learning (CS).
%, and (iii) an expert in program synthesis (Expert). 
% We also include example sets from a program synthesis expert (Expert) to demonstrate the concepts are learnable by a program synthesis system.
% \textcolor{blue}{JL (instead of the last sentence part): We also include a minimal example set provided by an expert in program synthesis, to demonstrate that the problems are learnable by the program synthesis system (Expert).}

As mentioned in the introduction, machine learning researchers often optimise their algorithms using synthetic datasets.
We investigate whether human-provided examples differ from random/synthetic examples.
Therefore, our third question is:
\begin{description}
\item[Q3.] Do humans provide better examples than randomly sampled examples?  
\end{description}
\noindent
To answer \textbf{Q3}, we compare the learning performance of synthesis systems trained on human-provided examples versus randomly generated ones.

\subsection{Method}

\subsubsection{List Concepts}
We use six list manipulation concepts
shown in Table \ref{tab:verbal_description}.
These concepts are commonly taught in introductory computer science courses and are standard benchmarks for evaluating both inductive program synthesis systems \cite{hofmann2009unifying,dreamcoder,popper,dagstuhl2024} and human learning \cite{rulehacker}. 
% \jl{@Celine We mentioned this both in the introduction and related work. Maybe a bit redundant?}
We selected these concepts because they are accessible to non-experts with no programming background.
% Figure \ref{fig:dropk} shows an example program for the concept \emph{dropk}.
 % The technical appendix includes an example of programs for each of these concepts.
% Each concept has a deterministic solution with a worst-case time complexity of $O(n)$ \ac{see discord}.
For each of these concepts, recursion is necessary to write a program that generalises to lists of arbitrary length.
% \ac{why mention complexity? is it relevant?} \ch{yes, it shows they are not too complicated. Reviewer 1 was interested in it}
Although a program implementing each of these concepts could theoretically operate on lists with elements of any type, we instruct participants to use natural numbers between 0 and 100.

\begin{table*}[ht]
% \small
\centering
\begin{footnotesize}
% \begin{tabular}{@{}l|l|p{11cm}}
% % \toprule
% \textbf{Name} & \textbf{Concept} & \textbf{Description}\\\midrule
% \emph{last} & \emph{Find the last element of a list} & Given as input a list, return the last element of that list.\\
% \emph{length} & \emph{Find the length of a list} & Given as input a list, return the number of elements in that list.\\
% \emph{append} & \emph{Append an element to a list} & Given as input a list and a natural number, return the list with the number inserted at the end of the list.\\
% \emph{maxlist} & \emph{Find the maximum element of a list} & Given as input a list, return the element with the highest value in that list.\\
% \emph{dropk} & \emph{Drop the first k elements of a list} & Given as input a list and a natural number k, return the input list without its first k elements.\\
% \emph{sorted} & \emph{Identify a sorted list} & Given as input a list, return TRUE if the list is sorted in ascending order, and FALSE if it is not.\\
% \end{tabular}
\begin{tabular}{@{}l|p{15cm}}
% \toprule
\textbf{Concept} & \textbf{Description}\\\midrule
\emph{last} & Given as input a list, return the last element of that list.\\
\emph{length} & Given as input a list, return the number of elements in that list.\\
\emph{append} & Given as input a list and a natural number, return the list with the number inserted at the end of the list.\\
\emph{maxlist} & Given as input a list, return the element with the highest value in that list.\\
\emph{dropk} & Given as input a list and a natural number $k$, return the input list without its first $k$ elements.\\
\emph{sorted} & Given as input a list, return \emph{true} if the list is sorted in ascending order, and \emph{false} if it is not.\\
\end{tabular}
\end{footnotesize}
\caption{
Textual descriptions of the programming concepts evaluated.
}
\label{tab:verbal_description}
\end{table*}

\subsubsection{Participants}
% \ch{explain how we recruited participants?}
% \ac{@Ute, what do you think? it is important to include such information for this type of study}\us{@all: Yes, I would give a bit more details, I include some text}
We recruited participants by e-mail. We invited (i) individuals without a computer science background, and (ii) computer science students (BSc, MSc, and PhD).
A total of 39 participants completed the study: 14 declared having no computer science background (NCS group), and 25 declared having a computer science background (CS group). 
%\textcolor{blue}{JL: I would not name the Expert group here as we will not be performing any statistical analyses, and this would highlight its benchmark position.}
% The mean age of participants is 25 $\pm$ 5.
The NCS group spent an average of 36min $\pm$ 8min answering the survey and the CS group 21min $\pm$ 3min.
Additionally, a postgraduate researcher with expertise in program synthesis (expert) participated in the study. This individual knows which program synthesis systems are tested and is familiar with their underlying algorithms. We use their examples as a baseline.
%The expert is familiar with the underlying algorithm of the program synthesis system used. 
% One CS participant provided identical examples and we decided to exclude their answers.
% \ch{which participants do we exclude, if any, and why?}
% \ch{do we need any other demographic information? highest degree of education?}
% \ac{@Ute, what do you think?}
% \us{I think highest degree of education is not necessary to report here, however, it might be of interest that the CS group is all students but the NCS group not.}
% The NCS group spent 36min18s $\pm$ 27min47s answering the survey and the CS group 20min52s $\pm$ 16min0s.
% The technical appendix shows the time spent to provide example sets for each of the tasks. 
% \ch{is this information helpful in the appendix?}
% \ac{@Ute, what do you think?}
% \us{I think, a technical appendix is not necessary here. However, we should have a documentation for our own reference where all details of the data are given.}

\subsubsection{Data collection}
% \ac{Should this part be under the examples section?}
We collect data through a web-based survey application. We first show participants the following introduction:
\begin{quotation}
\noindent
\emph{This study is on teaching concepts to computers by providing examples. The study focuses on list manipulation concepts.} 
\end{quotation}
% We then present the definition of a list.
% \begin{quotation}
% \noindent
% \emph{A list is a sequence of comma-separated values between square brackets, such as [1,2,3] or [7,4,15,8]. An empty list is written as [].}
% \end{quotation}
We then ask participants to provide examples in the form of input-output pairs to teach the given concepts:
\begin{quotation}
\noindent
\emph{We will give you a verbal description of a concept. We will then ask you to provide examples of the concept that you think are necessary to teach the concept to a computer. An example has both an input and an output. Inputs and outputs can be lists, natural numbers, Booleans (TRUE/FALSE) or ‘none’ (no value). Other symbols (negative integers, fractions, …) are disallowed. Use only natural numbers between 0 and 100. A concept might have one or two inputs. 
}
\end{quotation}
We show a worked-out example task (\emph{‘count the number of even numbers in a list’}). This example task includes a textual description of the concept (\emph{‘Given as input a list, find the number of even numbers in that list.’}) together with three input-output examples ($[0,2,4,6,8] \mapsto 5$; $[9,7,5,3,1] \mapsto 0$; and $[0,5,9,4,3,1,6,7,8,10,2] \mapsto 6$).
We tell participants that they will see several verbal descriptions of concepts and give them the following instructions:
\begin{quotation}
\noindent 
\emph{You can enter up to ten (10) examples for each concept. Try to explain the concept using as few examples as you think are necessary to teach the concept to a computer and give the examples in the most informative order.}
\end{quotation}
Next, we present participants with the six textual descriptions listed in Table~\ref{tab:verbal_description}, in a randomised order. 
% \ch{@JL is it true that the order of tasks was randomised?} \jl{@CH yes, they were randomised! :)}
Finally, we ask participants to provide demographic information, including their computer science background.

The technical appendix includes the full set of instructions presented to participants. The instructions do not mention program synthesis, as non-expert participants have little to no experience with it. Participants do not interact with the learning systems nor receive feedback while providing examples.

% \textbf{Illegal symbols and formatting errors.} 
\subsubsection{Data Cleaning}
Some participants included \emph{invalid symbols} in their examples. 
The number of examples with invalid symbols from the NCS and CS groups respectively are $N_\text{NCS} = 0$ and $N_\text{CS} = 5$.
For instance, a participant provided the example $[a,b,c,d,e,f,g] \mapsto 7$ for \emph{length}, which is invalid because only natural numbers are allowed as elements.
We correct invalid symbols by replacing them with the nearest admissible symbol. However, we do not correct elements greater than 100, as the synthesis systems used in the study can handle them.
% \ac{do we have a justification why?}.
Additionally, some participants provided examples in the \emph{wrong format} ($N_\text{NCS} = 13$, $N_\text{CS} = 5$). 
For instance, a participant wrote a semicolon instead of a comma. 
% Another participant provided the input integer with the output for \emph{dropk}. 
We re-formatted these examples. 
We discuss other types of errors, left uncorrected, in Section \ref{errors}.
% \ac{I think this paragraph should be part of the method}

\subsubsection{Random examples} 
To determine whether human-provided examples lead to better performance than randomly generated ones, we compare human-provided examples against examples drawn from two different random distributions:

\textbf{RandomUniform.} We sample the length of lists in examples from a uniform distribution between 0 and 100, and list elements from a uniform distribution between 0 and 100.

\textbf{RandomNormal.} We hypothesise that longer examples might be more informative, leading to better performance. 
To test this hypothesis, we generate examples in which the length of lists follows a normal distribution that reproduces the length distribution of examples provided by non-experts. 
We truncate this distribution to ensure positive lengths. We determine its parameters (mean, standard deviation, minimum, and maximum) from the participants' examples for each concept.
% \ac{@CH, can you clarify why a ``Truncated normal distribution'' and how it is truncated?}
% For both distributions, 
% \ac{I think this paragraph should be part of the method}
% \subsubsection{Reproducibility} The data and code appendix includes the anonymised data collected from the NCS group, the CS group, and the expert, the programs learned by \popper{}, as well as the code to reproduce our results.

\subsubsection{Learning Systems} 
% \ac{@CH, this section needs work. It would be good to write a sentence of two about how the systems work, to show they use a breadth of techniques}
% We use the following program synthesis systems.
% \textbf{\popper{}.} \popper{} \cite{popper,combo} is a state-of-the-art inductive logic programming system.
% \textbf{\metagol}
% \textbf{\ale{}.}
We use a portfolio of program synthesis systems and select the best-performing one for each set of examples. 
% We use three inductive logic programming (ILP) \cite{mugg:ilp,DBLP:journals/jair/CropperD22} systems, they learn Prolog programs.
We use the following systems.

\textbf{\popper{}.}
\popper{} \cite{popper,combo} is an ILP \cite{mugg:ilp} system based on constraint programming. 
\popper{} tolerates noisy examples and can learn recursive programs from only positive examples. \popper{} is guaranteed to learn a solution (a program that correctly generalises the training examples) if one exists in the search space.
It has been used on list function tasks \cite{decomp}.

\textbf{\metagol{}.}
\metagol{} \cite{mugg:metagold} is an ILP system based on a Prolog meta-interpreter. It also can learn recursive programs from positive examples only and is guaranteed to learn a solution.
It has also been used on list function tasks \cite{metabias,ruleefficient}.

\textbf{\ale{}.}
\ale{} \cite{aleph} is a separate-and-conquer ILP system which tolerates noisy examples. 

\textbf{\textsc{hl}.} Hacker-Like (\textsc{hl}) \cite{rulehacker} is a Bayesian inductive learning system designed to model human-like learning. It uses Monte Carlo tree search to search over metaprograms. It is specifically designed to learn list manipulation programs.
% and iteratively refines hypotheses through hacker mechanisms.

\textbf{\textsc{DeepSeek-Coder}.} 
\textsc{DeepSeek-Coder-V2} \cite{zhu2024deepseek} is a Mixture-of-Experts code language model trained on a variety of programming languages. We use the small 16B parameter version hosted locally within the ollama framework. To align with the ILP systems used, we prompt it to generate Prolog programs. However, since Prolog is not explicitly supported by the model, we also separately instruct it to generate Python programs, assuming that this language is properly represented in the training data due to its widespread usage. We evaluate the performance of the Prolog and Python programs independently and use the best-performing one.
%We generate Python and Prolog programs and use the best performing one.\ch{@JL shall we mention anything else?}
% \ac{explain what it is}

\subsubsection{Learning Settings}
We use the built-in bias for \textsc{hl}. Following the literature \cite{popper}, we allow the other systems to learn programs with the relations \emph{max}, \emph{head}, \emph{tail}, \emph{decrement}, \emph{increment}, \emph{geq}, \emph{eq}, \emph{empty}, \emph{zero}, \emph{one}, \emph{even}, and \emph{odd}. We also use the relation \emph{comps} for the concept \emph{append}.

For \textsc{DeepSeek-Coder}, we use a two-stage prompting process. In the first prompt, we provide the training examples, together with the arity and types of the target concept. We instruct the system to explain the function that produces the given input-output pairs and to generate corresponding code (in either Prolog or Python). Since the system's responses do not follow a consistent format, we use a second prompt to extract and return only the code from the previous response. We found that this approach yields more reliable and higher-quality results than directly prompting for code alone.
Once a program is generated, we test it against the training examples. If the program contains syntax errors or fails on any examples, we re-prompt the model with the errors and request a revised program. 
We allow up to three revisions, as we observed that the model rarely improves its program beyond three revisions.
The appendix shows our prompts.

\subsubsection{Experimental Procedure}
For each programming concept and each participant $\pi$, we train each synthesis system $s$ separately using the first $n$ examples provided by $\pi$. 
We evaluate the predictive accuracy (the proportion of correct predictions on unseen test data) of the program learned by $s$ using $2000$ test examples.
% \ac{positive or negative test examples?}
The default accuracy is 50\%. 
We aggregate the predictive accuracies across all synthesis systems and report the maximum. In other words, we report the highest accuracy achieved by any synthesis system to evaluate whether at least one synthesis system can effectively learn from the examples provided.
% \ac{In other words, .... [I will think how to rephrase to make it clearer].}
We vary the number of training examples $n$ from 1 to 10 and repeat training. 
We repeat these steps for each participant and programming concept.
We compare the predictive accuracies of the NCS and CS groups against those of the expert and the random distributions (\emph{RandomUniform} and \emph{RandomNormal}). We calculate the mean and standard deviation. Error bars in the figures and tables denote standard deviation. 
We also evaluate the statistics of the examples provided by participants. Box plots display the minimum, first quartile, median, third quartile, and maximum values. We use an Intel compute node with dual 2.0 GHz Intel Xeon Gold 6138 processors, 40 CPU cores, and 192 GB of DDR4 memory.

% \subsubsection{Ethics statement} 
% No ethics statement was required by the home university for this empirical study. %\jl{I removed the redundant footnote on page 1}
% For the conducted empirical study, no ethics statement have been required by the home university.
% \ch{TODO} \ch{@Ute, do you have any idea whether we should have an ethics statement?} \us{I am a bit unsure whether we should have taken care of this. There is no standard requirement for an ethics statement at German universities. Papers at CogSci do not include an ethics statement. I am also unsure about the wording, maybe "For the conducted empirical study, no ethics statement have been required by the home university."}

\subsection{Results} \label{sec:results}

We present the results of our empirical study.

\subsubsection{Q1. Do humans provide sufficient examples?}

Figures \ref{fig:acc_naive} and \ref{fig:acc_intermediate} show the average predictive accuracies of the best-performing synthesis system trained on the first $n$ examples provided by each participant, aggregated across all programming concepts.
% Note that the accuracies of programs learned by the LLM are much lower than the ones learned by \popper{}, with a much higher standard deviation \ac{Where does the figure show this result?}.
% This \ac{result?} is partly due to the LLM often not learning a syntactically correct program for some example sets, despite being informed about the errors in the refinement prompts \ac{refinement prompt is not mentioned elsehwere. What is it?} 
% Due to the unpredictable performance of the programs learned by the LLM, we refer to the programs learned by \popper{} for the statistical evaluation.
% \ac{does average = mean?} \ch{yes}
The appendix shows the accuracies for each programming concept.
These figures show that the expert consistently provides sufficient examples, leading at least one synthesis system to achieve maximal predictive accuracy (100\%) for every programming concept.  
This result demonstrates the existence of a set of examples from which a synthesis system can learn each target concept. 
Since this is an existential statement, we only need one expert.
% \ac{the writing in this paragraph needs a big rewrite.}

% We compare expert examples with those from the NCS group (Figure \ref{fig:acc_naive}) and CS group (Figure \ref{fig:acc_intermediate}).
% Examples from both the NCS and CS groups result in lower accuracy than the Expert on average. 
% \ac{[can we replace the two sentences above with this one?] 
By contrast, Figures \ref{fig:acc_naive} and \ref{fig:acc_intermediate} show that both the NCS and CS groups achieve lower accuracy than the expert on average.
This result suggests that non-experts often struggle to provide examples from which a program synthesis system correctly identifies the desired concept.
In other words, humans can provide sufficient examples but non-experts often do not. For instance, for the concept \emph{sorted}, none of the NCS participants achieve maximal accuracy. 
% In particular, no NCS participant included the base case (the empty list) in their examples.
Out of 14 NCS participants, 11 achieve at least 70\% accuracy, and 7 at least 90\%.
Out of 25 CS participants, only 13 achieve at least 70\% accuracy, 10 at least 90\%, and 3 maximum (100\%) accuracy.
% \ac{@CH, if there is space, can we mention on specific task where there is a large difference betweene expert and NCS?}

Overall, these results suggest that the answer to \textbf{Q1} is \textbf{no}: non-expert humans generally struggle to provide sufficient examples to teach programming concepts to synthesis systems. 
We discuss potential reasons for this result in Section \ref{sec:discuss}.

\begin{figure}[h]
    \centering
\begin{tikzpicture}[scale=0.7]
    \begin{axis}[
    xlabel=Number of examples,
    ylabel=Accuracy (\%),
    xmin=1,
    xmax=10,
    ymin=50,
    height=6cm,width=10cm,
    xtick={2,4,6,8,10},
     ytick={50, 60, 70, 80, 90, 100},
    ylabel style={yshift=-3mm},
    label style={font=\Large},
    legend style={legend pos=south east,font=\normalsize,style={nodes={right}}}
    ]

\addplot+[blue!70,mark=square*,error bars/.cd,y fixed,y dir=both,y explicit] coordinates {
(1, 60.19523809523809) +- (0, 1.9242437880745522)
(2, 77.56785714285715) +- (0, 2.546486285102101)
(3, 85.50059523809523) +- (0, 2.231746814167561)
(4, 88.5172619047619) +- (0, 1.9802326511264328)
(5, 90.08690476190476) +- (0, 1.9095397039941227)
(6, 90.93214285714285) +- (0, 1.8265185845598848)
(7, 91.49761904761905) +- (0, 1.7884211364186497)
(8, 92.56428571428573) +- (0, 1.6738116864925978)
(9, 92.4690476190476) +- (0, 1.6700154047393607)
(10, 92.76309523809525) +- (0, 1.6660092189477147)
};

\addplot+[black!30!green,mark=triangle*,mark options={scale=1.5},dashed,error bars/.cd,y fixed,y dir=both,y explicit] coordinates {
(1, 89.82692307692307) +- (0, 2.0855639125720855)
(2, 92.83076923076922) +- (0, 1.6946491838876983)
(3, 94.08525641025639) +- (0, 1.4349330035767796)
(4, 95.13205128205126) +- (0, 1.2346870830724839)
(5, 96.02628205128208) +- (0, 1.1180659039610372)
(6, 95.6570512820513) +- (0, 1.174303672189592)
(7, 97.10705128205129) +- (0, 1.0350302649277918)
(8, 96.95961538461538) +- (0, 0.8756031448148831)
(9, 96.94038461538462) +- (0, 0.9076689109425847)
(10, 97.35897435897436) +- (0, 0.8341895752229878)
};

\addplot+[black!70,mark=diamond*,mark options={scale=1.5},dashed,error bars/.cd,y fixed,y dir=both,y explicit] coordinates {
(1, 80.0673076923077) +- (0, 2.6119456532771803)
(2, 90.05641025641026) +- (0, 2.1011664376594616)
(3, 93.13910256410259) +- (0, 1.6645564348313664)
(4, 94.6352564102564) +- (0, 1.475803880506569)
(5, 96.799358974359) +- (0, 1.0913253808578576)
(6, 96.97564102564104) +- (0, 1.0558164477250922)
(7, 97.30000000000001) +- (0, 1.0111890673871675)
(8, 98.86538461538464) +- (0, 0.5341371115699602)
(9, 99.23205128205129) +- (0, 0.5660311422926304)
(10, 98.44166666666669) +- (0, 0.7596303765077518)
};

\addplot+[red!70,mark=otimes*,error bars/.cd,y fixed,y dir=both,y explicit] coordinates {
(1, 91.06666666666668) +- (0, 8.223722460729759)
(2, 99.88333333333334) +- (0, 0.07817359599705613)
(3, 99.94166666666668) +- (0, 0.04901813723284256)
(4, 100.0) +- (0, 0.0)
(5, 100.0) +- (0, 0.0)
(6, 100.0) +- (0, 0.0)
(7, 100.0) +- (0, 0.0)
(8, 100.0) +- (0, 0.0)
(9, 100.0) +- (0, 0.0)
(10, 100.0) +- (0, 0.0)
};

% \addplot+[cyan!70,mark=pentagon*,error bars/.cd,y fixed,y dir=both, y explicit] coordinates {
% (1,60.97) +- (0,21.82)
% (2,58.39) +- (0,27.68)
% (3,65.33) +- (0,23.98)
% (4,68.81) +- (0,25.01)
% (5,64.28) +- (0,29.17)
% (6,64.76) +- (0,27.72)
% (7,67.62) +- (0,26.16)
% (8,66.21) +- (0,26.53)
% (9,66.8) +- (0,26.44)
% (10,65.99) +- (0,25.94)
% };
    \legend{NCS participants, RandomUniform, RandomNormal, expert, LLM}
    \end{axis}
  \end{tikzpicture}
    \caption{
    Predictive accuracies for the NCS group averaged over all concepts when trained on progressively larger example sets.
    }
    \label{fig:acc_naive}
\end{figure}

\begin{figure}[h]
    \centering
\begin{tikzpicture}[scale=.7]
    \begin{axis}[
    xlabel=Number of examples,
    ylabel=Accuracy (\%),
    xmin=1,
    xmax=10,
    ymin=50,
    height=6cm,width=10cm,
    xtick={2,4,6,8,10},
     ytick={50, 60, 70, 80, 90, 100},
    ylabel style={yshift=-3mm},
    label style={font=\Large},
    legend style={legend pos=south east,font=\normalsize,style={nodes={right}}}
    ]
\addplot+[blue!70,mark=square*,error bars/.cd,y fixed,y dir=both,y explicit] coordinates {
(1, 58.15433333333334) +- (0, 1.4544530046991717)
(2, 66.87533333333332) +- (0, 1.8609940421069713)
(3, 75.872) +- (0, 1.9658824023035562)
(4, 80.09266666666667) +- (0, 1.9060126449826367)
(5, 81.00566666666667) +- (0, 1.8917773053021008)
(6, 83.35866666666666) +- (0, 1.8070412705903274)
(7, 83.20533333333333) +- (0, 1.8123208788551706)
(8, 83.34233333333334) +- (0, 1.7955964490261747)
(9, 84.10233333333333) +- (0, 1.7910634073109202)
(10, 83.54933333333332) +- (0, 1.8030350987950707)
};

\addplot+[black!30!green,mark=triangle*,mark options={scale=1.5},dashed,error bars/.cd,y fixed,y dir=both,y explicit] coordinates {
(1, 82.89233333333334) +- (0, 1.8154640686307142)
(2, 91.029) +- (0, 1.388911687478802)
(3, 92.98700000000001) +- (0, 1.1651301575859803)
(4, 92.70733333333332) +- (0, 1.2301389844056287)
(5, 92.95466666666667) +- (0, 1.1764279101966404)
(6, 91.99166666666667) +- (0, 1.2841867713396748)
(7, 93.52566666666667) +- (0, 1.1516446499253645)
(8, 96.104) +- (0, 0.8491706073327913)
(9, 96.18733333333331) +- (0, 0.8741236821996371)
(10, 95.51300000000002) +- (0, 0.9117832209786302)
};

\addplot+[black!70,mark=diamond*,mark options={scale=1.5},dashed,error bars/.cd,y fixed,y dir=both,y explicit] coordinates {
(1, 75.21200000000002) +- (0, 1.9601421374750214)
(2, 85.24133333333333) +- (0, 1.7471549738388956)
(3, 88.66199999999999) +- (0, 1.5658790213869294)
(4, 88.55199999999999) +- (0, 1.595374457597732)
(5, 91.70366666666666) +- (0, 1.304290896228136)
(6, 92.86966666666669) +- (0, 1.2985594235288178)
(7, 94.74100000000001) +- (0, 1.1371407360970212)
(8, 94.22366666666666) +- (0, 1.2124168088037826)
(9, 96.55833333333334) +- (0, 0.9032170215782059)
(10, 97.04666666666665) +- (0, 0.8578141497931729)

};

\addplot+[red!70,mark=otimes*,error bars/.cd,y fixed,y dir=both,y explicit] coordinates {
(1, 91.06666666666668) +- (0, 8.223722460729759)
(2, 99.88333333333334) +- (0, 0.07817359599705613)
(3, 99.94166666666668) +- (0, 0.04901813723284256)
(4, 100.0) +- (0, 0.0)
(5, 100.0) +- (0, 0.0)
(6, 100.0) +- (0, 0.0)
(7, 100.0) +- (0, 0.0)
(8, 100.0) +- (0, 0.0)
(9, 100.0) +- (0, 0.0)
(10, 100.0) +- (0, 0.0)
};

% \addplot+[cyan!70,mark=pentagon*,error bars/.cd,y fixed,y dir=both, y explicit] coordinates {
% (1,59.13) +- (0,19.89)
% (2,61.0) +- (0,22.74)
% (3,63.94) +- (0,27.04)
% (4,64.04) +- (0,26.31)
% (5,64.9) +- (0,24.97)
% (6,65.0) +- (0,25.66)
% (7,63.83) +- (0,26.04)
% (8,63.91) +- (0,24.63)
% (9,64.06) +- (0,24.96)
% (10,64.37) +- (0,24.7)
% };

    \legend{CS participants, RandomUniform, RandomNormal, expert, LLM}
    \end{axis}
  \end{tikzpicture}
    \caption{
    Predictive accuracies for the CS group averaged over all concepts when trained on progressively larger example sets.
    }
    \label{fig:acc_intermediate}
\end{figure}

\subsubsection{Q2. Does having a background in computer science improve a human's teaching ability?}

% \ch{@JL can you describe the statistical test used?}
% A mixed ANOVA, with group as the between-subject variable and task as the within-subject variable, shows there is no significant main effect for the factor group ($p=.10$). \ch{@JL do you know a reference for Anova?}
Table \ref{tab:max_acc} shows the predictive accuracies for each concept achieved from all of the examples provided by an NCS or CS participant.
% \ac{the results show that .... ?} 
% To analyse these results, we conduct a statistical test. 
 % \ac{so what? why is this sentence here?}. 
% Most example sets result in either maximum accuracy (100\%) or default accuracy (50\%), and few sets achieve intermediate values \ac{unsure why this sentence is here.}.
% \jl{For statistics, there are too many references. @Me Remove any footnotes and additional explanations, move normality tests to the main text as a quick mention and cite each test only once.}
% \ac{I agree}
A single Mann-Whitney U-test \cite{mann1947test}
%\footnote{The Mann-Whitney U-test is a non-parametric alternative to the Student's t-test. We use this test since a Shapiro-Wilk test \cite{shapiro1965analysis} indicates that the accuracies are not normally distributed (most example sets result in either maximum accuracy (100\%) or default accuracy (50\%), and few sets achieve intermediate values).}
 finds no statistically significant differences between the NCS and CS groups for any of the concepts. We use a non-parametric test because a Shapiro-Wilk test \cite{shapiro1965analysis} indicates that the accuracies are not normally distributed (most example sets result in either maximum accuracy (100\%) or default accuracy (50\%) and only a few sets achieve intermediate values).
% \ac{is this sentence the most important one. Should it not come above where I put 'the results show ... '?}
This result suggests that in this context and among non-experts a computer science background does not significantly affect the ability to teach a programming concept to a synthesis system.
% The lack of a significant difference suggests that, in this context and among non-experts, the background of participants does not enhance their ability to teach a concept to a program synthesis system.
In other words, domain-specific knowledge alone may not always enhance teaching effectiveness. 

However, the expert consistently achieves maximal accuracy (100\%), which may indicate that understanding the learner’s algorithm plays an important role.
This observation suggests that effective teaching might instead require both familiarity with the concepts and an understanding of how the learner’s algorithm works. Further investigation is needed to confirm this hypothesis and clarify the role of algorithmic understanding in teaching effectiveness.

Overall, these results suggest that the answer to \textbf{Q2} is likely \textbf{no}: a participant's background does not significantly impact their ability to teach a programming concept to a synthesis system.
However, understanding the learner’s algorithm could play an important role in effective teaching.
% While understanding of the domain might not be beneficial, comprehension of the learner's learning algorithm can be.
% An understanding of the domain does not help, but an understanding of the learner works can help. 
% Only when the participants understand how a learner works. 
% \jl{This feels as if the lack of specificity in the research question was used to answer this question positively. The research question we tested for was rather "Does a background in computer science influence the human's ability to teach...?" I would be more comfortable with a "No" to Q2, with mention of the observation that it is not understanding of the algorithms that matter, but understanding of the learner.}

\begin{table}[h]
% \small
\centering
\begin{footnotesize}
% \footnotesize
\begin{tabular}{@{}l|cc@{}}
% \toprule
& \textbf{NCS} & \textbf{CS} \\\midrule
% \multirow{2}{*}{\textbf{Concept}} & \multicolumn{2}{c}{\textbf{NCS}} & \multicolumn{2}{c}{\textbf{CS}} \\
% & \textit{mean} & \textit{sd} & \textit{mean} & \textit{sd} \\ \midrule
\emph{last} & $89 \pm 21$ & $\boldsymbol{93} \pm  17$ \\
\emph{length} & $\boldsymbol{100} \pm 0$ & $94 \pm 16$ \\
\emph{append} & $\boldsymbol{95} \pm 11$ & $\boldsymbol{95} \pm 14$ \\
\emph{maxlist} & $\boldsymbol{96} \pm 13$ & $79 \pm 24$ \\
\emph{dropk} & $\boldsymbol{92} \pm 18$ & $66 \pm 23$ \\
\emph{sorted} & $\boldsymbol{85} \pm 17$ & $74 \pm 19$ \\
% \textbf{Concept} & \textbf{NCS} & \textbf{CS}\\\midrule
% \emph{last} & 89 $\pm$ 6  & \textbf{93 $\pm$ 3}\\
% \emph{length} & \textbf{100 $\pm$ 0} & 94 $\pm$ 3\\
% \emph{append} & \textbf{95 $\pm$ 3} & \textbf{95 $\pm$ 3}\\
% \emph{maxlist} & \textbf{96 $\pm$ 3} & 79 $\pm$ 5\\
% \emph{dropk} & \textbf{92 $\pm$ 5} & 66 $\pm$ 5\\
% \emph{sorted} & \textbf{96 $\pm$ 3} & 74 $\pm$ 4\\

% \textbf{Task} & \textbf{NCS} & \textbf{CS} & \textbf{Expert}\\\midrule
% \emph{last} & 89 $\pm$ 6  & 95 $\pm$ 3 & \textbf{100}\\
% \emph{length} & \textbf{100 $\pm$ 0} & 96 $\pm$ 3 & \textbf{100}\\
% \emph{append} & 95 $\pm$ 3 & 92 $\pm$ 3 & \textbf{100}\\
% \emph{maxlist} & 96 $\pm$ 4 & 82 $\pm$ 5 & \textbf{100}\\
% \emph{dropk} & 92 $\pm$ 5 & 80 $\pm$ 5 & \textbf{100}\\
% \emph{sorted} & 84 $\pm$ 5 & 77 $\pm$ 4 & \textbf{100}\\
\end{tabular}
\end{footnotesize}
\caption{
Mean and standard deviation of predictive accuracies for full example sets for the NCS and CS groups.
% \ch{@JL, is the difference statistically significant for some of the tasks? If so, we could add stars for these tasks in this table.}
% \jl{Report $n$. Ute and I think we should not report percentages but use fractions (as usual in Psychology and ML evaluations). We should report Mean and standard deviation here, remove the $\pm$, make subcolumns for mean and sd instead.}
% The error bars denote standard error.
% \ac{Would a box plot (or some other plot) be easier to interpret than this table?}
% \ch{is the boxplot below easier to understand?}
% \ch{shall we show similar figures / tables for N = 2, 5, ... examples?}
}
\label{tab:max_acc}
\end{table}

\subsubsection{Q3. Do humans provide better examples than randomly sampled examples?}
Figures \ref{fig:acc_naive} and \ref{fig:acc_intermediate} 
show the average predictive accuracies of the best-performing synthesis systems trained with non-expert examples and examples with length sampled from a uniform or normal distribution. 
% A mixed ANOVA confirms the significance of the difference between the humans and the random distributions for each of the CS and NCS groups ($p < 0.1$ \ch{@JL was it the threshold used?}).
% Non-expert human participants did not provide better examples than randomly sampled examples.
% To account for the truncated normal distribution's parameter dependency on population metrics, we perform multiple statistical tests.
For each participant group (CS or NCS), we conduct tests using the predictive accuracies from two, three, or all ten examples, resulting in a total of six analyses. 
We choose these specific points to test how informative the first examples are, and the full sets of examples.
% have one test using the full data we collected, but also take the initial examples into account since we assume that humans conscientiously choose examples that are more appropriate than sampling randomly.
% \ac{why is this sentence important? It is needed to motivate the significance test?}.
A Kruskal-Wallis H-test \cite{kruskal1952use}, %\footnote{The Kruskal-Wallis H-test is a non-parametric variant of the independent one-way ANOVA. We use this test because, similarly to participant data, accuracies from randomly generated examples do not follow a normal distribution.}
adjusted with a Benjamini-Hochberg false discovery control method \cite{benjamini1995controlling} to account for family-wise error rate\footnote{When performing multiple statistical tests on the same data, the probability of not falsely rejecting a null hypothesis is $(1-\alpha)^n$ with $n$ being the number of tests performed.}, shows a significant group effect when using two examples for the NCS group and two or ten examples for the CS group.
% \jl{H-test only shows that there is some significant group effect. We cannot infer from it between which groups the difference is, this needs the U-test. Also, BH does not correct the test, but adjusts p-values when multiple tests are conducted.}
Post-hoc U-tests %\cite{mann1947test} 
further identify that, for the NCS group, accuracies from human-provided examples are significantly lower than accuracies from the uniform distribution when using 2 examples ($p=.01$). For the CS group, accuracies from human-provided examples are significantly lower than those from both random distributions when using 2 examples ($p<.001$), and lower than accuracies from the normal distribution when using 10 examples.
These results indicate that the NCS and CS groups perform worse than both random distributions. We discuss potential reasons for these findings in Section \ref{sec:discuss}.

% the uniform distribution when using 10 examples a significant difference between the CS group and the normal distribution when using 10 examples ($p<.01$), and both random distributions when using 2 examples ($p<.001$). These results show that, compared to both a uniform and parameter-matching normal distribution, the NCS and CS groups have worse performance. We discuss potential reasons for this result in Section \ref{sec:discuss}.

Overall, these results suggest that the answer to \textbf{Q3} is likely \textbf{no}: non-expert humans do not provide better examples than randomly sampled ones.

\section{Discussion}
\label{sec:discuss}

The results from Section \ref{sec:results} suggest that non-experts, regardless of their background in computer science, struggle to provide sufficient examples to teach programming concepts to synthesis systems.
In this section, we discuss potential explanations.

\subsubsection{Erroneous Examples}
\label{errors}

One potential explanation for the lower accuracy of the NCS and CS groups is the presence of errors in their examples. We identify two types of errors. 

A \emph{systematic error} occurs when all examples in an example set contain the same error ($N_\text{NCS} = 3$, $N_\text{CS} = 8$).
This type of error suggests that the participant provided examples for a different concept. 
For instance, an NCS participant provided examples for the concept \emph{``return the first k elements of a list''} for \emph{dropk}, such as $[4,8,7,6,2,1,5,4,8,3,1],2 \mapsto [4,8]$, when the correct output is $[7,6,2,1,5,4,8,3,1]$. 
Similarly, a CS participant provided examples for the concept \emph{``remove the element k in the input list''} for \emph{dropk} and provided examples such as $[20,13,42,53,23,21],23 \mapsto [20,13,42,53,21]$.

A \emph{non-systematic error} is an error on a single example of an example set ($N_\text{NCS} = 1$, $N_\text{CS} = 3$). For instance, a CS participant provided the example $[10,0,59,68,23,42,53] \mapsto 59$ for \emph{maxlist}, and an NCS participant gave $[99999,7676768] \mapsto false$ for \emph{sorted}.

To evaluate the impact of these errors, we exclude examples with either a systematic or non-systematic error and retrain the synthesis systems. 
Table \ref{tab:max_acc_corrected} shows the resulting predictive accuracies. %\ch{@JL is there a statistical difference between the results in Figure 2 and 3? if not we should mention it}
%\ch{@JL can you please also check the LLM on the corrected training sets?}
A U-test reveals a significant difference between the NCS and CS groups for \textit{dropk}, with the NCS group performing better. This result suggests that, in some cases, humans with domain-specific knowledge may provide less suitable examples than those without. However, examples provided by humans without such knowledge are more impacted by errors.
% \jl{This formulation feels a bit clunky, maybe someone else has more luck with it than me :)}

While excluding erroneous examples improves accuracy, they still are not maximal. This result suggests that, even without errors, the examples provided by non-experts are insufficient for the systems to correctly generalise. For instance, for \emph{sorted}, a participant provided only the examples $[1,2,3,4,5] \mapsto true; [0,0,0,0,1] \mapsto true; 
[5,4,3,2,1]\mapsto false; [1,0,0,0,0] \mapsto false$, from which a system incorrectly learns that a list is sorted if its second element is greater than or equal to its first.
% \ac{some past / present tenses in this paragraph. Rephrase to be consistent.}

In summary, this result suggests that, while human errors limit performance, participants also did not provide sufficient examples.

\begin{table}[h]
% \small
\centering
\begin{footnotesize}
% \footnotesize
\begin{tabular}{@{}l|cc@{}}
% \toprule

% \multirow{2}{*}{\textbf{Concept}} & \multicolumn{2}{c}{\textbf{NCS}} & \multicolumn{2}{c}{\textbf{CS}} \\
& \textbf{NCS} & \textbf{CS} \\\midrule
% & \textit{mean} & \textit{sd} & \textit{mean} & \textit{sd} \\ \midrule
\emph{last} & $89 \pm 21$ & $\boldsymbol{93} \pm  17$ \\
\emph{length} & $\boldsymbol{100} \pm 0$ & $94 \pm 16$ \\
\emph{append} & $95 \pm 11$ & $\boldsymbol{100} \pm 0$ \\
\emph{maxlist} & $\boldsymbol{96} \pm 13$ & $79 \pm 25$ \\
\emph{dropk}$^*$ & $\boldsymbol{99} \pm 1$ & $69 \pm 25$ \\
\emph{sorted} & $\boldsymbol{84} \pm 17$ & $74 \pm 19$ \\

% \textbf{Concept} & \textbf{NCS} & \textbf{CS}\\\midrule
% \emph{last} & 92 $\pm$ 5  & \textbf{95 $\pm$ 3}\\
% \emph{length} & \textbf{100 $\pm$ 0} & 96 $\pm$ 3\\
% \emph{append} & 95 $\pm$ 3 & \textbf{98 $\pm$ 2}\\
% \emph{maxlist} & \textbf{96 $\pm$ 4} & 81 $\pm$ 5\\
% \emph{dropk} & \textbf{99 $\pm$ 0} & 89 $\pm$ 5\\
% \emph{sorted} & \textbf{96 $\pm$ 3} & 77 $\pm$ 4\\

% \begin{tabular}{@{}l|ccc@{}}
% % \toprule
% \textbf{Task} & \textbf{NCS} & \textbf{CS} & \textbf{Expert}\\\midrule
% \emph{last} & 92 $\pm$ 5  & 95 $\pm$ 3 & \textbf{100}\\
% \emph{length} & \textbf{100 $\pm$ 0} & 96 $\pm$ 3 & \textbf{100}\\
% \emph{append} & 95 $\pm$ 3 & 98 $\pm$ 2 & \textbf{100}\\
% \emph{maxlist} & 96 $\pm$ 4 & 81 $\pm$ 5 & \textbf{100}\\
% \emph{dropk} & 99 $\pm$ 0 & 89 $\pm$ 5 & \textbf{100}\\
% \emph{sorted} & 83 $\pm$ 5 & 77 $\pm$ 4 & \textbf{100}\\
\end{tabular}
\end{footnotesize}
\caption{
Mean and standard deviation of predictive accuracies for full example sets for the  NCS and CS groups, after ignoring examples with systematic or non-systematic errors. $^*:p<.05$
% The error bars denote standard error.
% \ch{would a box plot be better to show this information? or a curve versus number of examples?}
% \ch{I simply ignored example sets with systematic or non-systematic errors}
% \jl{There is no definition of the errors before this graph. Could move this to Discussion, show that there is very little impact.}
}
\label{tab:max_acc_corrected}
\end{table}

% \ch{which of these differences are statistically significant?} \jl{Compared to not excluding the errors? I could check, but I am 100\% certain none of them.}
% \ch{not sure this explanation is helpful if results are not significant}
% sorted
% pos(f([1,2,3,4,5,6,7,8,9])).
% pos(f([1,3,5,7,9])).
% neg(f([5,7,1,3,6,8,9])).
% \ch{shall we run Popper without these errors to see their impact?}

\subsubsection{Simpler Examples}
Another potential factor that may contribute to lower accuracy is the simplicity of non-experts' examples.
To examine this explanation, we analyse the number of examples, list lengths, and element values provided by participants.

\paragraph{Number of examples.} Figure \ref{fig:n_exs} shows the distribution of the number of examples provided by participants.
It shows both the NCS and CS groups typically provided more examples than the expert for all concepts.
% \ch{is the difference statistically significant? do we need a test in this situation?}
% \ac{Ideally, yes, but as it is not a major claim, I think we do not need a significance test}
For instance, the expert provided only 3 examples for \emph{maxlist}, while the NCS and CS groups provided $6.7 \pm 0.7$ and $6.6 \pm 0.5$ examples, respectively.
% \ac{no need for two decimal places, just say 6 and 7, or 6.1 and 6.}
% \ac{Also need to mention some measure of variance}
% \ch{@JL, can you please check whether the difference is statistically significant, and if so write something about the test used?}
% A Shapiro-Wilk test \cite{shapiro1965analysis} indicates that the number of examples is not normally distributed \ac{does this matter?}.
An H-test %\cite{kruskal1952use}
reveals a significant group effect ($p<.05$). Post-hoc U-tests % \cite{mann1947test} 
show that the expert provided significantly fewer examples than both the NCS and CS groups across all concepts ($p<.05$). We use a non-parametric test since a Shapiro-Wilk test %\cite{shapiro1965analysis}
indicates that the number of examples is not normally distributed. 
This result might surprise the reader, as one might expect that more examples would improve teaching effectiveness.
However, as \citeauthor{telle2019teaching} [\citeyear{telle2019teaching}] note, effective machine teaching depends not only on the number of examples but also on their length.
% it is crucial to consider not only the number of examples but also the length of examples in example sets.

\input{fig_number_exs}

\paragraph{Length of lists.} Figure \ref{fig:list_length} shows the distribution of list lengths of the examples provided by participants. Both NCS and CS participants generally provide examples of similar lengths across programming concepts, whereas the expert adjusts their example lengths based on the concept.
For instance, for \emph{last} and \emph{length}, at least 75\% of examples have fewer than 6 elements for non-experts (both NCS and CS) and the expert.
However, for \emph{maxlist}, the expert provided examples of length 0, 12, and 14, while at least 75\% of examples have fewer than 7 elements for non-experts (NCS and CS).
% This concept is challenging for non-experts, who often do not achieve maximal accuracy.
This difference is important because longer examples can contain more bits and hence carry more information.
For instance, for \emph{last}, a non-expert provided two examples of length 4, leading a synthesis system to learn the concept \emph{return the fourth element of the input list}, which is simpler to express compared to the concept of \emph{last}. By contrast, the expert provided a single example of length 6. A synthesis system learns the concept of \emph{last} because the concept \emph{return the sixth element of the list} is more complex to express compared to \emph{last}.
Non-experts may provide shorter examples because they are easier to generate and cognitively simpler.
The expert likely includes longer examples based on their deeper understanding of the synthesis systems. 
% Figures \ref{fig:acc_naive} and \ref{fig:acc_intermediate} show that sampling list lengths randomly can improve learning performance. This results suggests that the length of lists greatly influences the ability to teach a synthesis system a concept.
% \ac{this paragraph needs breaking down as it is contains many topics}

% Moreover, they might assume that the learner processes information similarly to a human, leading them to avoid complex lists as they are more difficult to parse. 
% \ch{need to link better with these results}
% according to a uniform distribution bounded by 1 and 100 .
% However, non-expert humans provided shorter examples. 
% Therefore, we also compare the performance of human participants against a normal distribution with mean and standard deviation inferred from the examples of the human participants for each task.
% Figures \ref{fig:acc_naive} and \ref{fig:acc_intermediate} show the normal distribution leads to performance closer to both the NCS and CS groups. 
% \ac{are these figures supposed to be different from those in the previous section? they refer to the same figures}

\input{fig_exs_lengths}
% Another explanation for why the NCS and CS groups did not achieve maximal accuracy is their examples do not carry enough information.
% One explanation for this result is that randomly generated examples carry more information. 

\paragraph{Element values.} Figure \ref{fig:element_values} shows the distribution of element values in the lists provided by participants. It shows that elements in lists from both non-experts tend to have smaller values and less variability compared to those provided by the expert. For instance, for \emph{sorted}, 75\% of elements are below 9 for the CS group and below 20 for the NCS group. By contrast, 50\% of elements in the expert’s examples range between 8 and 43.
This difference is important because lower variability in element values increases the likelihood of coincidental patterns.
For instance, for \emph{sorted}, an NCS participant provides the examples $[1,3,6,9,14,18]\mapsto true;
[1,2,3,4,5] \mapsto true; [0,1] \mapsto true; [3,1,12,2,7] \mapsto false$. From these examples, a synthesis system incorrectly learns that a list is sorted if its first element is strictly less than 2.
By contrast, the expert provides more diverse examples and uses sorted lists with unique, non-repeating elements.

\input{fig_exs_values}

% By contrast, the sorted lists provided by the expert contain unique elements, with no element repeated in another sorted list.
% \ac{can we add something about the expert? ``by contrast, the expert  .... ''}
% \ac{what is this random paragraph doing here? I think it should be at the start of the previous paragraph}

Overall, these results suggest that the quality of examples, rather than their quantity, plays a crucial role in effective teaching. 

%% file: fig_number_exs.tex
\definecolor{mygreen}{rgb}{0.3,0.6,0.4}

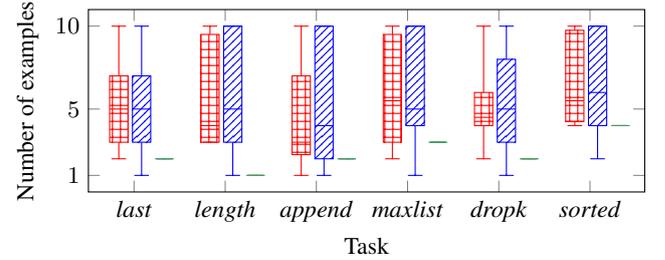
\begin{figure}[h]
\begin{minipage}{0.5\textwidth}
\begin{tikzpicture}
\begin{customlegend}[legend columns=3,legend style={nodes={scale=1, transform shape},align=left,column sep=0ex},
        legend entries={Expert, NCS, CS}]
        \addlegendimage{red,mark=square*,fill,fill opacity=0.2}
        \addlegendimage{mygreen,mark=square*,
        pattern=grid,pattern color=mygreen}
        % \addlegendimage{\node[pattern=dots]}

        \addlegendimage{blue,mark=square*,pattern=north east lines, pattern color=blue}
\end{customlegend}
\end{tikzpicture}
\end{minipage}\hfill\\
\begin{minipage}{0.5\textwidth}
\begin{tikzpicture}
    \node[rotate=90] {
\begin{tikzpicture}
\begin{axis}[
xmin=0, xmax=11,
ymin=-1, ymax=23,
height=8.2cm,
width=4cm,
xticklabel style={font=\footnotesize, rotate=270},
yticklabel style={font=\footnotesize, rotate=-90},
xlabel style={font=\footnotesize},
xticklabel pos=top,
xlabel=Number of exs,
% x={(0cm,0.11cm)}, y={(01cm,0cm)},
ytick={1,5,9,13,17,21},
yticklabels=
{\emph{dropk},\emph{sorted},\emph{append},\emph{maxlist},\emph{length},\emph{last}},
cycle list={{mygreen},{blue},{red}},
boxplot={
        draw position={4*floor(\plotnumofactualtype/3) + mod(\plotnumofactualtype,3)},
        % box extend=0.2
}
]

\addplot+[
pattern=grid,
pattern color=mygreen,
    boxplot prepared={
      lower whisker=2,
      lower quartile=4.0,
      median=4.5,
      upper quartile=6.0,
      upper whisker=10,
    },
    ] coordinates {};
\addplot+[
pattern=north east lines,
pattern color=blue,
    boxplot prepared={
      lower whisker=1,
      lower quartile=3.0,
      median=5.0,
      upper quartile=8.0,
      upper whisker=10,
    },
    ] coordinates {};
\addplot+[
fill,
line width=2pt,
    boxplot prepared={
      lower whisker=2,
      lower quartile=2.0,
      median=2.0,
      upper quartile=2.0,
      upper whisker=2,
    },
    ] coordinates {};

\addplot+[
pattern=grid,
pattern color=mygreen,
    boxplot prepared={
      lower whisker=4,
      lower quartile=4.25,
      median=5.5,
      upper quartile=9.75,
      upper whisker=10,
    },
    ] coordinates {};
\addplot+[
pattern=north east lines,
pattern color=blue,
    boxplot prepared={
      lower whisker=2,
      lower quartile=4.0,
      median=6.0,
      upper quartile=10.0,
      upper whisker=10,
    },
    ] coordinates {};
\addplot+[
fill,
line width=2pt,
    boxplot prepared={
      lower whisker=4,
      lower quartile=4.0,
      median=4.0,
      upper quartile=4.0,
      upper whisker=4,
    },
    ] coordinates {};

\addplot+[
pattern=grid,
pattern color=mygreen,
    boxplot prepared={
      lower whisker=1,
      lower quartile=2.25,
      median=3.0,
      upper quartile=7.0,
      upper whisker=10,
    },
    ] coordinates {};
\addplot+[
pattern=north east lines,
pattern color=blue,
    boxplot prepared={
      lower whisker=1,
      lower quartile=2.0,
      median=4.0,
      upper quartile=10.0,
      upper whisker=10,
    },
    ] coordinates {};
\addplot+[
fill,
line width=2pt,
    boxplot prepared={
      lower whisker=2,
      lower quartile=2.0,
      median=2.0,
      upper quartile=2.0,
      upper whisker=2,
    },
    ] coordinates {};

\addplot+[
pattern=grid,
pattern color=mygreen,
    boxplot prepared={
      lower whisker=2,
      lower quartile=3.0,
      median=5.5,
      upper quartile=9.5,
      upper whisker=10,
    },
    ] coordinates {};
\addplot+[
pattern=north east lines,
pattern color=blue,
    boxplot prepared={
      lower whisker=1,
      lower quartile=4.0,
      median=5.0,
      upper quartile=10.0,
      upper whisker=10,
    },
    ] coordinates {};
\addplot+[
fill,
line width=2pt,
    boxplot prepared={
      lower whisker=3,
      lower quartile=3.0,
      median=3.0,
      upper quartile=3.0,
      upper whisker=3,
    },
    ] coordinates {};

\addplot+[
pattern=grid,
pattern color=mygreen,
    boxplot prepared={
      lower whisker=3,
      lower quartile=3.0,
      median=4.0,
      upper quartile=9.5,
      upper whisker=10,
    },
    ] coordinates {};
\addplot+[
pattern=north east lines,
pattern color=blue,
    boxplot prepared={
      lower whisker=1,
      lower quartile=3.0,
      median=5.0,
      upper quartile=10.0,
      upper whisker=10,
    },
    ] coordinates {};
\addplot+[
fill,
line width=2pt,
    boxplot prepared={
      lower whisker=1,
      lower quartile=1.0,
      median=1.0,
      upper quartile=1.0,
      upper whisker=1,
    },
    ] coordinates {};

\addplot+[
pattern=grid,
pattern color=mygreen,
    boxplot prepared={
      lower whisker=2,
      lower quartile=3.0,
      median=5.0,
      upper quartile=7.0,
      upper whisker=10,
    },
    ] coordinates {};
\addplot+[
pattern=north east lines,
pattern color=blue,
    boxplot prepared={
      lower whisker=1,
      lower quartile=3.0,
      median=5.0,
      upper quartile=7.0,
      upper whisker=10,
    },
    ] coordinates {};
\addplot+[
fill,
line width=2pt,
    boxplot prepared={
      lower whisker=2,
      lower quartile=2.0,
      median=2.0,
      upper quartile=2.0,
      upper whisker=2,
    },
    ] coordinates {};

\addplot[
mark=*,
only marks,
point meta =explicit symbolic,
fill opacity=0.01, mark size=1.2pt,
]
table[x=data,y=label,meta=label]{"last_nex_expert.dat"};

% \addplot[
% mark=|,
% only marks,
% point meta =explicit symbolic,
% mark size=3pt,
% ]
% table{"last_nex_expert_av.dat"};

\addplot[
mark=*,
only marks,
point meta =explicit symbolic,
fill opacity=0.01, mark size=1.2pt,
]
table[x=data,y=label,meta=label]{"last_nex_ncs.dat"};

% \addplot[
% mark=|,
% only marks,
% point meta =explicit symbolic,
% mark size=3pt,
% ]
% table{"last_nex_ncs_av.dat"};

\addplot[
mark=*,
only marks,
point meta =explicit symbolic,
fill opacity=0.01, mark size=1.2pt,
]
table[x=data,y=label,meta=label]{"last_nex_cs.dat"};

% \addplot[
% mark=|,
% only marks,
% point meta =explicit symbolic,
% mark size=3pt,
% ]
% table{"last_nex_cs_av.dat"};

\addplot[
mark=*,
only marks,
point meta =explicit symbolic,
fill opacity=0.01, mark size=1.2pt,
]
table[x=data,y=label,meta=label]{"length_nex_expert.dat"};

% \addplot[
% mark=|,
% only marks,
% point meta =explicit symbolic,
% mark size=3pt,
% ]
% table{"length_nex_expert_av.dat"};

\addplot[
mark=*,
only marks,
point meta =explicit symbolic,
fill opacity=0.01, mark size=1.2pt,
]
table[x=data,y=label,meta=label]{"length_nex_ncs.dat"};

% \addplot[
% mark=|,
% only marks,
% point meta =explicit symbolic,
% mark size=3pt,
% ]
% table{"length_nex_ncs_av.dat"};

\addplot[
mark=*,
only marks,
point meta =explicit symbolic,
fill opacity=0.01, mark size=1.2pt,
]
table[x=data,y=label,meta=label]{"length_nex_cs.dat"};

% \addplot[
% mark=|,
% only marks,
% point meta =explicit symbolic,
% mark size=3pt,
% ]
% table{"length_nex_cs_av.dat"};

\addplot[
mark=*,
only marks,
point meta =explicit symbolic,
fill opacity=0.01, mark size=1.2pt,
]
table[x=data,y=label,meta=label]{"maxlist_nex_expert.dat"};

% \addplot[
% mark=|,
% only marks,
% point meta =explicit symbolic,
% mark size=3pt,
% ]
% table{"maxlist_nex_expert_av.dat"};

\addplot[
mark=*,
only marks,
point meta =explicit symbolic,
fill opacity=0.01, mark size=1.2pt,
]
table[x=data,y=label,meta=label]{"maxlist_nex_ncs.dat"};

% \addplot[
% mark=|,
% only marks,
% point meta =explicit symbolic,
% mark size=3pt,
% ]
% table{"maxlist_nex_ncs_av.dat"};

\addplot[
mark=*,
only marks,
point meta =explicit symbolic,
fill opacity=0.01, mark size=1.2pt,
]
table[x=data,y=label,meta=label]{"maxlist_nex_cs.dat"};

% \addplot[
% mark=|,
% only marks,
% point meta =explicit symbolic,
% mark size=3pt,
% ]
% table{"maxlist_nex_cs_av.dat"};

\addplot[
mark=*,
only marks,
point meta =explicit symbolic,
fill opacity=0.01, mark size=1.2pt,
]
table[x=data,y=label,meta=label]{"append_nex_expert.dat"};

% \addplot[
% mark=|,
% only marks,
% point meta =explicit symbolic,
% mark size=3pt,
% ]
% table{"append_nex_expert_av.dat"};

\addplot[
mark=*,
only marks,
point meta =explicit symbolic,
fill opacity=0.01, mark size=1.2pt,
]
table[x=data,y=label,meta=label]{"append_nex_ncs.dat"};

% \addplot[
% mark=|,
% only marks,
% point meta =explicit symbolic,
% mark size=3pt,
% ]
% table{"append_nex_ncs_av.dat"};

\addplot[
mark=*,
only marks,
point meta =explicit symbolic,
fill opacity=0.01, mark size=1.2pt,
]
table[x=data,y=label,meta=label]{"append_nex_cs.dat"};

% \addplot[
% mark=|,
% only marks,
% point meta =explicit symbolic,
% mark size=3pt,
% ]
% table{"append_nex_cs_av.dat"};

\addplot[
mark=*,
only marks,
point meta =explicit symbolic,
fill opacity=0.01, mark size=1.2pt,
]
table[x=data,y=label,meta=label]{"sorted_nex_expert.dat"};

% \addplot[
% mark=|,
% only marks,
% point meta =explicit symbolic,
% mark size=3pt,
% ]
% table{"sorted_nex_expert_av.dat"};

\addplot[
mark=*,
only marks,
point meta =explicit symbolic,
fill opacity=0.01, mark size=1.2pt,
]
table[x=data,y=label,meta=label]{"sorted_nex_ncs.dat"};

% \addplot[
% mark=|,
% only marks,
% point meta =explicit symbolic,
% mark size=3pt,
% ]
% table{"sorted_nex_ncs_av.dat"};

\addplot[
mark=*,
only marks,
point meta =explicit symbolic,
fill opacity=0.01, mark size=1.2pt,
]
table[x=data,y=label,meta=label]{"sorted_nex_cs.dat"};

% \addplot[
% mark=|,
% only marks,
% point meta =explicit symbolic,
% mark size=3pt,
% ]
% table{"sorted_nex_cs_av.dat"};

\addplot[
mark=*,
only marks,
point meta =explicit symbolic,
fill opacity=0.01, mark size=1.2pt,
]
table[x=data,y=label,meta=label]{"dropk_nex_expert.dat"};

% \addplot[
% mark=|,
% only marks,
% point meta =explicit symbolic,
% mark size=3pt,
% ]
% table{"dropk_nex_expert_av.dat"};

\addplot[
mark=*,
only marks,
point meta =explicit symbolic,
fill opacity=0.01, mark size=1.2pt,
]
table[x=data,y=label,meta=label]{"dropk_nex_ncs.dat"};

% \addplot[
% mark=|,
% only marks,
% point meta =explicit symbolic,
% mark size=3pt,
% ]
% table{"dropk_nex_ncs_av.dat"};

\addplot[
mark=*,
only marks,
point meta =explicit symbolic,
fill opacity=0.01, mark size=1.2pt,
]
table[x=data,y=label,meta=label]{"dropk_nex_cs.dat"};

% \addplot[
% mark=|,
% only marks,
% point meta =explicit symbolic,
% mark size=3pt,
% ]
% table{"dropk_nex_cs_av.dat"};

\end{axis}
\end{tikzpicture}
    };
    \end{tikzpicture}
\end{minipage}
\caption{Number of examples provided by participants.\label{fig:n_exs}
}
\end{figure}

%% file: fig_exs_lengths.tex
\begin{figure}[h]
\begin{minipage}{0.5\textwidth}
\begin{tikzpicture}
\begin{customlegend}[legend columns=3,legend style={nodes={scale=1, transform shape},align=left,column sep=0ex},
        legend entries={Expert, NCS, CS}]
        \addlegendimage{red,mark=square*,fill,fill opacity=0.2}
        \addlegendimage{mygreen,mark=square*,
        pattern=grid,pattern color=mygreen}
        % \addlegendimage{\node[pattern=dots]}

        \addlegendimage{blue,mark=square*,pattern=north east lines, pattern color=blue}
\end{customlegend}
\end{tikzpicture}
\end{minipage}\hfill\\
\begin{minipage}{0.5\textwidth}
\begin{tikzpicture}
    \node[rotate=90] {
\begin{tikzpicture}
\begin{axis}[
xmin=-1, xmax=22,
ymin=-1, ymax=23,
height=8.2cm,
width=4cm,
xticklabel style={font=\footnotesize, rotate=270},
yticklabel style={font=\footnotesize, rotate=-90},
xlabel style={font=\footnotesize},
xticklabel pos=top,
xlabel=Example lengths,
% x={(0cm,0.11cm)}, y={(01cm,0cm)},
ytick={1,5,9,13,17,21},
yticklabels=
{\emph{dropk},\emph{sorted},\emph{append},\emph{maxlist},\emph{length},\emph{last}},
cycle list={{mygreen},{blue},{red}},
boxplot={
        draw position={4*floor(\plotnumofactualtype/3) + mod(\plotnumofactualtype,3)},
        % box extend=0.2
}
]

% DROPK
\addplot+[
pattern=grid,
pattern color=mygreen,
    boxplot prepared={
      lower whisker=0,
      lower quartile=4.0,
      median=5.0,
      upper quartile=7.0,
      upper whisker=12,
    },
    ] coordinates {};
\addplot+[
pattern=north east lines,
pattern color=blue,
    boxplot prepared={
      lower whisker=0,
      lower quartile=4.0,
      median=5.0,
      upper quartile=7.0,
      upper whisker=19,
    },
    ] coordinates {};
\addplot+[
fill,fill opacity=0.2,
    boxplot prepared={
      lower whisker=3,
      lower quartile=4.5,
      median=6.0,
      upper quartile=7.5,
      upper whisker=9,
    },
    ] coordinates {};

% SORTED
\addplot+[
pattern=grid,
pattern color=mygreen,
    boxplot prepared={
      lower whisker=1,
      lower quartile=4.0,
      median=5.0,
      upper quartile=6.0,
      upper whisker=10,
    },
    ] coordinates {};
\addplot+[
pattern=north east lines,
pattern color=blue,
    boxplot prepared={
      lower whisker=0,
      lower quartile=3.0,
      median=5.0,
      upper quartile=6.0,
      upper whisker=10,
    },
    ] coordinates {};
\addplot+[
fill,fill opacity=0.2,
    boxplot prepared={
      lower whisker=0,
      lower quartile=6.0,
      median=8.0,
      upper quartile=8.0,
      upper whisker=8,
    },
    ] coordinates {};

% APPEND
\addplot+[
pattern=grid,
pattern color=mygreen,
    boxplot prepared={
      lower whisker=0,
      lower quartile=3.0,
      median=4.0,
      upper quartile=4.25,
      upper whisker=10,
    },
    ] coordinates {};

\addplot+[
pattern=north east lines,
pattern color=blue,
    boxplot prepared={
      lower whisker=0,
      lower quartile=3.0,
      median=4.0,
      upper quartile=5.0,
      upper whisker=10,
    },
    ] coordinates {};
\addplot+[
fill,fill opacity=0.2,
    boxplot prepared={
      lower whisker=3,
      lower quartile=4.75,
      median=6.5,
      upper quartile=8.25,
      upper whisker=10,
    },
    ] coordinates {};
% MAXLIST
\addplot+[
pattern=grid,
pattern color=mygreen,
    boxplot prepared={
      lower whisker=1,
      lower quartile=4.0,
      median=5.0,
      upper quartile=6.0,
      upper whisker=12,
    },
    ] coordinates {};
\addplot+[
pattern=north east lines,
pattern color=blue,
    boxplot prepared={
      lower whisker=0,
      lower quartile=3.0,
      median=5.0,
      upper quartile=7.0,
      upper whisker=14,
    },
    ] coordinates {};
\addplot+[
fill,fill opacity=0.2,
    boxplot prepared={
      lower whisker=0,
      lower quartile=6.0,
      median=12.0,
      upper quartile=13.0,
      upper whisker=14,
    },
    ] coordinates {};

% LENGTH
\addplot+[
pattern=grid,
pattern color=mygreen,
    boxplot prepared={
      lower whisker=0,
      lower quartile=3.0,
      median=4.0,
      upper quartile=6.0,
      upper whisker=18,
    },
    ] coordinates {};
\addplot+[
pattern=north east lines,
pattern color=blue,
    boxplot prepared={
      lower whisker=0,
      lower quartile=1.0,
      median=4.0,
      upper quartile=6.0,
      upper whisker=12,
    },
    ] coordinates {};
\addplot+[
fill,fill opacity=0.2,
    boxplot prepared={
      lower whisker=4,
      lower quartile=4.0,
      median=4.0,
      upper quartile=4.0,
      upper whisker=4,
    },
    ] coordinates {};

% LAST
\addplot+[
pattern=grid,
pattern color=mygreen,
    boxplot prepared={
      lower whisker=1,
      lower quartile=3.0,
      median=5.0,
      upper quartile=6.0,
      upper whisker=14,
    },
    ] coordinates {};
\addplot+[
pattern=north east lines,
pattern color=blue,
    boxplot prepared={
      lower whisker=0,
      lower quartile=3.0,
      median=5.0,
      upper quartile=6.0,
      upper whisker=20,
    },
    ] coordinates {};
\addplot+[
fill,fill opacity=0.2,
    boxplot prepared={
      lower whisker=0,
      lower quartile=1.5,
      median=3.0,
      upper quartile=4.5,
      upper whisker=6,
    },
    ] coordinates {};
    
\addplot[
mark=*,
only marks,
point meta =explicit symbolic,
fill opacity=0.01, mark size=0.8pt,
]
table[x=data,y=label,meta=label]{"last_length_expert.dat"};

% \addplot[
% mark=|,
% only marks,
% point meta =explicit symbolic,
% mark size=3pt,
% ]
% table{"last_length_expert_av.dat"};

\addplot[
mark=*,
only marks,
point meta =explicit symbolic,
fill opacity=0.01, mark size=0.8pt,
]
table[x=data,y=label,meta=label]{"last_length_ncs.dat"};

% \addplot[
% mark=|,
% only marks,
% point meta =explicit symbolic,
% mark size=3pt,
% ]
% table{"last_length_ncs_av.dat"};

\addplot[
mark=*,
only marks,
point meta =explicit symbolic,
fill opacity=0.01, mark size=0.8pt,
]
table[x=data,y=label,meta=label]{"last_length_cs.dat"};

% \addplot[
% mark=|,
% only marks,
% point meta =explicit symbolic,
% mark size=3pt,
% ]
% table{"last_length_cs_av.dat"};

\addplot[
mark=*,
only marks,
point meta =explicit symbolic,
fill opacity=0.01, mark size=0.8pt,
]
table[x=data,y=label,meta=label]{"length_length_expert.dat"};

% \addplot[
% mark=|,
% only marks,
% point meta =explicit symbolic,
% mark size=3pt,
% ]
% table{"length_length_expert_av.dat"};

\addplot[
mark=*,
only marks,
point meta =explicit symbolic,
fill opacity=0.01, mark size=0.8pt,
]
table[x=data,y=label,meta=label]{"length_length_ncs.dat"};

% \addplot[
% mark=|,
% only marks,
% point meta =explicit symbolic,
% mark size=3pt,
% ]
% table{"length_length_ncs_av.dat"};

\addplot[
mark=*,
only marks,
point meta =explicit symbolic,
fill opacity=0.01, mark size=0.8pt,
]
table[x=data,y=label,meta=label]{"length_length_cs.dat"};

% \addplot[
% mark=|,
% only marks,
% point meta =explicit symbolic,
% mark size=3pt,
% ]
% table{"length_length_cs_av.dat"};

\addplot[
mark=*,
only marks,
point meta =explicit symbolic,
fill opacity=0.01, mark size=0.8pt,
]
table[x=data,y=label,meta=label]{"maxlist_length_expert.dat"};

% \addplot[
% mark=|,
% only marks,
% point meta =explicit symbolic,
% mark size=3pt,
% ]
% table{"maxlist_length_expert_av.dat"};

\addplot[
mark=*,
only marks,
point meta =explicit symbolic,
fill opacity=0.01, mark size=0.8pt,
]
table[x=data,y=label,meta=label]{"maxlist_length_ncs.dat"};

% \addplot[
% mark=|,
% only marks,
% point meta =explicit symbolic,
% mark size=3pt,
% ]
% table{"maxlist_length_ncs_av.dat"};

\addplot[
mark=*,
only marks,
point meta =explicit symbolic,
fill opacity=0.01, mark size=0.8pt,
]
table[x=data,y=label,meta=label]{"maxlist_length_cs.dat"};

% \addplot[
% mark=|,
% only marks,
% point meta =explicit symbolic,
% mark size=3pt,
% ]
% table{"maxlist_length_cs_av.dat"};

\addplot[
mark=*,
only marks,
point meta =explicit symbolic,
fill opacity=0.01, mark size=0.8pt,
]
table[x=data,y=label,meta=label]{"append_length_expert.dat"};

% \addplot[
% mark=|,
% only marks,
% point meta =explicit symbolic,
% mark size=3pt,
% ]
% table{"append_length_expert_av.dat"};

\addplot[
mark=*,
only marks,
point meta =explicit symbolic,
fill opacity=0.01, mark size=0.8pt,
]
table[x=data,y=label,meta=label]{"append_length_ncs.dat"};

% \addplot[
% mark=|,
% only marks,
% point meta =explicit symbolic,
% mark size=3pt,
% ]
% table{"append_length_ncs_av.dat"};

\addplot[
mark=*,
only marks,
point meta =explicit symbolic,
fill opacity=0.01, mark size=0.8pt,
]
table[x=data,y=label,meta=label]{"append_length_cs.dat"};

% \addplot[
% mark=|,
% only marks,
% point meta =explicit symbolic,
% mark size=3pt,
% ]
% table{"append_length_cs_av.dat"};

\addplot[
mark=*,
only marks,
point meta =explicit symbolic,
fill opacity=0.01, mark size=0.8pt,
]
table[x=data,y=label,meta=label]{"sorted_length_expert.dat"};

% \addplot[
% mark=|,
% only marks,
% point meta =explicit symbolic,
% mark size=3pt,
% ]
% table{"sorted_length_expert_av.dat"};

\addplot[
mark=*,
only marks,
point meta =explicit symbolic,
fill opacity=0.01, mark size=0.8pt,
]
table[x=data,y=label,meta=label]{"sorted_length_ncs.dat"};

% \addplot[
% mark=|,
% only marks,
% point meta =explicit symbolic,
% mark size=3pt,
% ]
% table{"sorted_length_ncs_av.dat"};

\addplot[
mark=*,
only marks,
point meta =explicit symbolic,
fill opacity=0.01, mark size=0.8pt,
]
table[x=data,y=label,meta=label]{"sorted_length_cs.dat"};

% \addplot[
% mark=|,
% only marks,
% point meta =explicit symbolic,
% mark size=3pt,
% ]
% table{"sorted_length_cs_av.dat"};

\addplot[
mark=*,
only marks,
point meta =explicit symbolic,
fill opacity=0.01, mark size=0.8pt,
]
table[x=data,y=label,meta=label]{"dropk_length_expert.dat"};

% \addplot[
% mark=|,
% only marks,
% point meta =explicit symbolic,
% mark size=3pt,
% ]
% table{"dropk_length_expert_av.dat"};

\addplot[
mark=*,
only marks,
point meta =explicit symbolic,
fill opacity=0.01, mark size=0.8pt,
]
table[x=data,y=label,meta=label]{"dropk_length_ncs.dat"};

% \addplot[
% mark=|,
% only marks,
% point meta =explicit symbolic,
% mark size=3pt,
% ]
% table{"dropk_length_ncs_av.dat"};

\addplot[
mark=*,
only marks,
point meta =explicit symbolic,
fill opacity=0.01, mark size=0.8pt,
]
table[x=data,y=label,meta=label]{"dropk_length_cs.dat"};

% \addplot[
% mark=|,
% only marks,
% point meta =explicit symbolic,
% mark size=3pt,
% ]
% table{"dropk_length_cs_av.dat"};

\end{axis}
\end{tikzpicture}
    };
    \end{tikzpicture}
\end{minipage}
\caption{Length of lists provided by participants.\label{fig:list_length}
}
\end{figure}
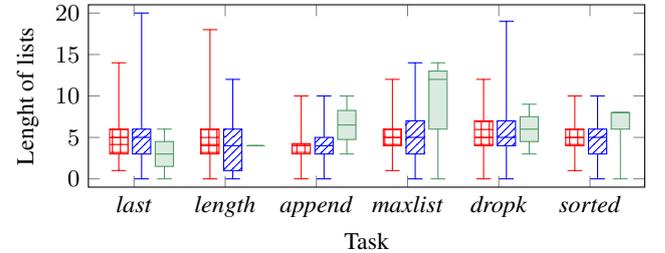

%% file: fig_exs_values.tex
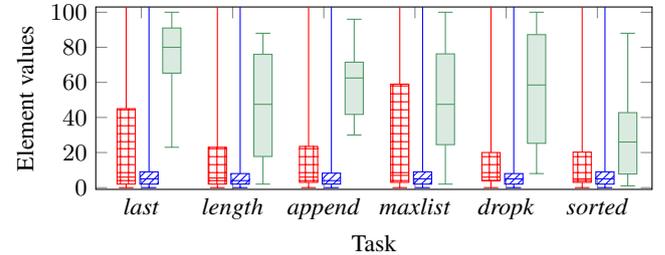
\begin{figure}[h]
\begin{minipage}{0.5\textwidth}
\begin{tikzpicture}
\begin{customlegend}[legend columns=3,legend style={nodes={scale=1, transform shape},align=left,column sep=0ex},
        legend entries={Expert, NCS, CS}]
        \addlegendimage{red,mark=square*,fill,fill opacity=0.2}
        \addlegendimage{mygreen,mark=square*,
        pattern=grid,pattern color=mygreen}
        % \addlegendimage{\node[pattern=dots]}

        \addlegendimage{blue,mark=square*,pattern=north east lines, pattern color=blue}
\end{customlegend}
\end{tikzpicture}
\end{minipage}\hfill\\
\begin{minipage}{0.5\textwidth}
\begin{tikzpicture}
    \node[rotate=90] {
\begin{tikzpicture}
\begin{axis}[
xmin=-5, xmax=130,
ymin=-1, ymax=23,
height=8.2cm,
width=4cm,
xticklabel style={font=\footnotesize, rotate=270},
yticklabel style={font=\footnotesize, rotate=-90},
xlabel style={font=\footnotesize},
xticklabel pos=top,
xlabel=Element values,
% x={(0cm,0.11cm)}, y={(01cm,0cm)},
ytick={1,5,9,13,17,21},
yticklabels=
{\emph{dropk},\emph{sorted},\emph{append},\emph{maxlist},\emph{length},\emph{last}},
cycle list={{mygreen},{blue},{red}},
boxplot={
        draw position={4*floor(\plotnumofactualtype/3) + mod(\plotnumofactualtype,3)},
        % box extend=0.2
}
]

% DROPK
\addplot+[
pattern=grid,
pattern color=mygreen,
    boxplot prepared={
      lower whisker=0,
      lower quartile=4.0,
      median=6.0,
      upper quartile=20.0,
      % upper whisker=83673,
    upper whisker=1000,
    },
    ] coordinates {};
\addplot+[
pattern=north east lines,
pattern color=blue,
    boxplot prepared={
      lower whisker=0,
      lower quartile=2.0,
      median=5.0,
      upper quartile=8.0,
      % upper whisker=20000,
    upper whisker=1000,
    },
    ] coordinates {};
\addplot+[
fill,fill opacity=0.2,
    boxplot prepared={
      lower whisker=8,
      lower quartile=25.25,
      median=58.5,
      upper quartile=87.25,
      upper whisker=100,
    },
    ] coordinates {};
    
% SORTED
\addplot+[
pattern=grid,
pattern color=mygreen,
    boxplot prepared={
      lower whisker=0,
      lower quartile=3.0,
      median=5.0,
      upper quartile=20.25,
      % upper whisker=5415151885,
    upper whisker=1000,
    },
    ] coordinates {};
\addplot+[
pattern=north east lines,
pattern color=blue,
    boxplot prepared={
      lower whisker=0,
      lower quartile=2.0,
      median=5.0,
      upper quartile=9.0,
      % upper whisker=1000000000000,
      upper whisker=1000,
    },
    ] coordinates {};
\addplot+[
fill,fill opacity=0.2,
    boxplot prepared={
      lower whisker=1,
      lower quartile=7.75,
      median=26.0,
      upper quartile=42.75,
      upper whisker=88,
    },
    ] coordinates {};

% APPEND
\addplot+[
pattern=grid,
pattern color=mygreen,
    boxplot prepared={
      lower whisker=0,
      lower quartile=3.0,
      median=6.0,
      upper quartile=23.5,
      % upper whisker=5641223,
        upper whisker=1000,
    },
    ] coordinates {};
\addplot+[
pattern=north east lines,
pattern color=blue,
    boxplot prepared={
      lower whisker=0,
      lower quartile=2.0,
      median=4.0,
      upper quartile=8.25,
      % upper whisker=6788872,
        upper whisker=1000,
    },
    ] coordinates {};
\addplot+[
fill,fill opacity=0.2,
    boxplot prepared={
      lower whisker=30,
      lower whisker=30,
      lower quartile=43.0,
      median=67.0,
      upper quartile=79.0,
      upper whisker=99,
    },
    ] coordinates {};
% MAXLIST
\addplot+[
pattern=grid,
pattern color=mygreen,
    boxplot prepared={
      lower whisker=0,
      lower quartile=3.0,
      median=7.0,
      upper quartile=59.0,
      % upper whisker=564541,
        upper whisker=1000,
    },
    ] coordinates {};
\addplot+[
pattern=north east lines,
pattern color=blue,
    boxplot prepared={
      lower whisker=0,
      lower quartile=2.0,
      median=5.0,
      upper quartile=9.0,
      % upper whisker=6321546,
        upper whisker=1000,
    },
    ] coordinates {};
\addplot+[
fill,fill opacity=0.2,
    boxplot prepared={
      lower whisker=2,
      lower quartile=24.5,
      median=47.5,
      upper quartile=76.25,
      upper whisker=100,
    },
    ] coordinates {};

% LENGTH
\addplot+[
pattern=grid,
pattern color=mygreen,
    boxplot prepared={
      lower whisker=0,
      lower quartile=2.0,
      median=5.5,
      upper quartile=23.0,
      % upper whisker=54452526,
    upper whisker=1000,
    },
    ] coordinates {};
\addplot+[
pattern=north east lines,
pattern color=blue,
    boxplot prepared={
      lower whisker=0,
      lower quartile=2.0,
      median=4.0,
      upper quartile=8.0,
      upper whisker=10000
    },
    ] coordinates {};
\addplot+[
fill,fill opacity=0.2,
    boxplot prepared={
      lower whisker=2,
      lower quartile=17.75,
      median=47.5,
      upper quartile=76.0,
      upper whisker=88,
    },
    ] coordinates {};

% LAST
\addplot+[
pattern=grid,
pattern color=mygreen,
    boxplot prepared={
      lower whisker=0,
      lower quartile=2.0,
      median=6.0,
      upper quartile=45.0,
      % upper whisker=786546,
    upper whisker=1000,
    },
    ] coordinates {};
\addplot+[
pattern=north east lines,
pattern color=blue,
    boxplot prepared={
      lower whisker=0,
      lower quartile=2.0,
      median=5.0,
      upper quartile=9.0,
      upper whisker=1000,
    },
    ] coordinates {};
\addplot+[
fill,fill opacity=0.2,
    boxplot prepared={
      lower whisker=23,
      lower quartile=65.25,
      median=80.0,
      upper quartile=91.0,
      upper whisker=100,
    },
    ] coordinates {};

\addplot[
mark=*,
only marks,
point meta =explicit symbolic,
fill opacity=0.003, mark size=0.5pt,
]
table[x=data,y=label,meta=label]{"last_values_expert.dat"};

% \addplot[
% mark=|,
% only marks,
% point meta =explicit symbolic,
% mark size=3pt,
% ]
% table{"last_values_expert_av.dat"};

\addplot[
mark=*,
only marks,
point meta =explicit symbolic,
fill opacity=0.003, mark size=0.5pt,
]
table[x=data,y=label,meta=label]{"last_values_ncs.dat"};

% \addplot[
% mark=|,
% only marks,
% point meta =explicit symbolic,
% mark size=3pt,
% ]
% table{"last_values_ncs_av.dat"};

\addplot[
mark=*,
only marks,
point meta =explicit symbolic,
fill opacity=0.003, mark size=0.5pt,
]
table[x=data,y=label,meta=label]{"last_values_cs.dat"};

% \addplot[
% mark=|,
% only marks,
% point meta =explicit symbolic,
% mark size=3pt,
% ]
% table{"last_values_cs_av.dat"};

\addplot[
mark=*,
only marks,
point meta =explicit symbolic,
fill opacity=0.003, mark size=0.5pt,
]
table[x=data,y=label,meta=label]{"length_values_expert.dat"};

% \addplot[
% mark=|,
% only marks,
% point meta =explicit symbolic,
% mark size=3pt,
% ]
% table{"length_values_expert_av.dat"};

\addplot[
mark=*,
only marks,
point meta =explicit symbolic,
fill opacity=0.003, mark size=0.5pt,
]
table[x=data,y=label,meta=label]{"length_values_ncs.dat"};

% \addplot[
% mark=|,
% only marks,
% point meta =explicit symbolic,
% mark size=3pt,
% ]
% table{"length_values_ncs_av.dat"};

\addplot[
mark=*,
only marks,
point meta =explicit symbolic,
fill opacity=0.003, mark size=0.5pt,
]
table[x=data,y=label,meta=label]{"length_values_cs.dat"};

% \addplot[
% mark=|,
% only marks,
% point meta =explicit symbolic,
% mark size=3pt,
% ]
% table{"length_values_cs_av.dat"};

\addplot[
mark=*,
only marks,
point meta =explicit symbolic,
fill opacity=0.003, mark size=0.5pt,
]
table[x=data,y=label,meta=label]{"maxlist_values_expert.dat"};

% \addplot[
% mark=|,
% only marks,
% point meta =explicit symbolic,
% mark size=3pt,
% ]
% table{"maxlist_values_expert_av.dat"};

\addplot[
mark=*,
only marks,
point meta =explicit symbolic,
fill opacity=0.003, mark size=0.5pt,
]
table[x=data,y=label,meta=label]{"maxlist_values_ncs.dat"};

% \addplot[
% mark=|,
% only marks,
% point meta =explicit symbolic,
% mark size=3pt,
% ]
% table{"maxlist_values_ncs_av.dat"};

\addplot[
mark=*,
only marks,
point meta =explicit symbolic,
fill opacity=0.003, mark size=0.5pt,
]
table[x=data,y=label,meta=label]{"maxlist_values_cs.dat"};

% \addplot[
% mark=|,
% only marks,
% point meta =explicit symbolic,
% mark size=3pt,
% ]
% table{"maxlist_values_cs_av.dat"};

\addplot[
mark=*,
only marks,
point meta =explicit symbolic,
fill opacity=0.003, mark size=0.5pt,
]
table[x=data,y=label,meta=label]{"append_values_expert.dat"};

% \addplot[
% mark=|,
% only marks,
% point meta =explicit symbolic,
% mark size=3pt,
% ]
% table{"append_values_expert_av.dat"};

\addplot[
mark=*,
only marks,
point meta =explicit symbolic,
fill opacity=0.003, mark size=0.5pt,
]
table[x=data,y=label,meta=label]{"append_values_ncs.dat"};

% \addplot[
% mark=|,
% only marks,
% point meta =explicit symbolic,
% mark size=3pt,
% ]
% table{"append_values_ncs_av.dat"};

\addplot[
mark=*,
only marks,
point meta =explicit symbolic,
fill opacity=0.003, mark size=0.5pt,
]
table[x=data,y=label,meta=label]{"append_values_cs.dat"};

% \addplot[
% mark=|,
% only marks,
% point meta =explicit symbolic,
% mark size=3pt,
% ]
% table{"append_values_cs_av.dat"};

\addplot[
mark=*,
only marks,
point meta =explicit symbolic,
fill opacity=0.003, mark size=0.5pt,
]
table[x=data,y=label,meta=label]{"sorted_values_expert.dat"};

% \addplot[
% mark=|,
% only marks,
% point meta =explicit symbolic,
% mark size=3pt,
% ]
% table{"sorted_values_expert_av.dat"};

\addplot[
mark=*,
only marks,
point meta =explicit symbolic,
fill opacity=0.003, mark size=0.5pt,
]
table[x=data,y=label,meta=label]{"sorted_values_ncs.dat"};

% \addplot[
% mark=|,
% only marks,
% point meta =explicit symbolic,
% mark size=3pt,
% ]
% table{"sorted_values_ncs_av.dat"};

\addplot[
mark=*,
only marks,
point meta =explicit symbolic,
fill opacity=0.003, mark size=0.5pt,
]
table[x=data,y=label,meta=label]{"sorted_values_cs.dat"};

% \addplot[
% mark=|,
% only marks,
% point meta =explicit symbolic,
% mark size=3pt,
% ]
% table{"sorted_values_cs_av.dat"};

\addplot[
mark=*,
only marks,
point meta =explicit symbolic,
fill opacity=0.003, mark size=0.5pt,
]
table[x=data,y=label,meta=label]{"dropk_values_expert.dat"};

% \addplot[
% mark=|,
% only marks,
% point meta =explicit symbolic,
% mark size=3pt,
% ]
% table{"dropk_values_expert_av.dat"};

\addplot[
mark=*,
only marks,
point meta =explicit symbolic,
fill opacity=0.003, mark size=0.5pt,
]
table[x=data,y=label,meta=label]{"dropk_values_ncs.dat"};

% \addplot[
% mark=|,
% only marks,
% point meta =explicit symbolic,
% mark size=3pt,
% ]
% table{"dropk_values_ncs_av.dat"};

\addplot[
mark=*,
only marks,
point meta =explicit symbolic,
fill opacity=0.003, mark size=0.5pt,
]
table[x=data,y=label,meta=label]{"dropk_values_cs.dat"};

% \addplot[
% mark=|,
% only marks,
% point meta =explicit symbolic,
% mark size=3pt,
% ]
% table{"dropk_values_cs_av.dat"};

\end{axis}
\end{tikzpicture}
    };
    \end{tikzpicture}
\caption{Values of elements provided by participants.\label{fig:element_values}
}
\end{minipage}
\end{figure}

%% file: 06-conclusions.tex
\section{Conclusions and Future Work}
%\ac{recap the main research question}
Our empirical study explores whether humans can teach machines to code. 
Our results are negative, and suggest that non-experts, regardless of their computer science background, struggle to provide sufficient examples to effectively teach list manipulation concepts to program synthesis systems. 
However, an expert could provide a small set of examples from which a synthesis system achieves perfect accuracy, showing that sufficient examples exist for the programming concepts tested.
Furthermore, our results suggest that synthesis systems perform better when trained on randomly generated examples than on those provided by non-expert humans.
% \ac{maybe reiterate that experts could provide sufficient examples}

% Our results show that it is not trivial to provide helpful examples to generalise the intended procedure for standard list manipulation problems.

\subsection{Limitations}
This study is the first to explore whether humans can teach machines to code. There are, therefore, limitations. 

% \ch{theoretical analysis from cognitive psychology etc?}

\noindent
\textbf{Eliciting teaching strategies.} We show participants a general instruction which did not specify that examples are for a program synthesis system, which may have led them to give general-purpose examples. However, our results show that CS participants made use of their knowledge and, for instance, often included an example for the base case for recursive concepts. Nevertheless, future work should explore how varying instructions to elicit specific teaching strategies, as proposed by \citeauthor{khan2011humans} [\citeyear{khan2011humans}] and \citeauthor{good_teaching} [\citeyear{good_teaching}], can help humans provide higher-quality examples.

\noindent
\textbf{Better systems.} We tested 5 synthesis systems, but none consistently learns the desired concepts from non-expert examples. 
One system consistently learns perfectly accurate programs from the expert examples.
However, an untested system might successfully learn from non-expert data. 
% It should also be noted that the experiments with LLM-based code generators were started before systems like OpenAI o1 \cite{jaech2024openai} or DeepSeek R1 \cite{guo2025deepseek} were available. With the resource investment of larger models, better program synthesis may be possible.
% Yet, this result highlights the need to adapt current systems to non-expert examples.
% Moreover, we purposely chose simple tasks that are understandable by non-experts. Future work could explore more complex tasks, such as tree manipulation tasks. Given our negative results with simple tasks, we anticipate similar negative results with more complex ones.
% \ac{why is more complex tasks part of the better systems limitation?}
% \ac{I suggest we cut the more complex tasks part}

\subsection{Outlook}
Our results have implications for the field of inductive program synthesis.
In PBE \cite{lieberman:pbe} and end-user programming \cite{gulwani2017program}, it is assumed that humans can provide sufficient examples to teach a concept.
Our results suggest this assumption does not always hold.
Furthermore, most of the program synthesis research relies on randomly generated examples for training.
Our results suggest that training using random examples yields substantially better performance compared to using non-expert examples. 
If these systems are intended to be eventually trained by humans, this result should motivate researchers to incorporate more human-generated examples into training datasets.

% Until now, prompting LLM-based code generators with input-output examples has not received research attention. Our results show that our preliminary implementation of this approach falls short of other known code generation tasks such as code generation from docstrings, code completion, or code correction \cite{codex,austin2021program}.
% \ch{I would remove this paragraph, we do not compare examples vs  other code generation, nor do these references}

Consequently, our study raises two challenges for the field.
First, our study emphasises the necessity to develop systems that better adapt to human examples, for instance with hybrid machine-human example selection.
Second, our results highlight the need to build algorithms capable of generating training sets of random examples  that more accurately reflect human-produced examples.

%% file: prompts.tex
\section{Prompts}
\label{sec:prompts}

For prompting \textsc{DeepSeek-Coder-V2}, we implement a simple message history that provides the context of previous prompts and responses per learning scenario. Prompts are dynamically composed for each task. In the following, we show two sample exchanges, one for Prolog and one for Python, using the expert examples for the task \textsl{last} (Table \ref{tab:expert_exs}).

\subsection{Prolog}

\textsl{\textbf{Prompt:} Your task is to learn a Prolog program from examples. Find the simplest program which covers all positive and no negative examples.} 
\noindent
\textsl{Here are examples of the Prolog predicate f/2 which takes a list as input and returns an element as output:}
\begin{verbatim}
pos(f([72,88,23,92,63,100],100)).
pos(f([],none)).
\end{verbatim}
\textsl{The predicates pos/1 and neg/1 mark positive and negative examples of the predicate f/2 respectively.\\
Please explain which predicate f/2 would result in the provided examples and why. Provide the Prolog definition for it as well.}
\vspace{5pt}

\noindent
\textsl{\textbf{Answer:}}
\textit{To find a simple Prolog program that covers all positive examples and no negative ones, we need to define a predicate `f/2` such that it matches the given examples correctly. Let's analyze the provided examples:
\begin{enumerate}
    \item The first example is \textit{pos(f([72,88,23,92,63,100],100))}, which means when the list \textit{[72, 88, 23, 92, 63, 100]} is given to `f/2`, it should return `100`.
    \item The second example is \textit{pos(f([],none))}, which means when an empty list is given to `f/2`, it should return `none`.
\end{enumerate}
From these examples, we can infer that:
\begin{itemize}
    \item If the list is not empty and contains the number 100 somewhere in its elements, then `f/2` should return `100`.
    \item If the list is empty, then `f/2` should return `none`.
\end{itemize}
Given this understanding, a possible definition for `f/2` could be:}
\begin{small}
\begin{verbatim}
% Base case: if the list is empty, 
% return 'none'
f([], none).
% Recursive case: check if the head 
% of the list is 100, and if so, 
% return it; otherwise, continue 
% with the tail.
f([H|T], H) :- H =:= 100, !. 
% The cut (!) operator ensures 
% that only the first match is 
% considered, preventing backtracking 
% into non-matching tails.
f([H|T], Result) :- H =\= 100, f(T, Result).
\end{verbatim}
\end{small}\vspace{2pt}
\textit{This definition covers both positive examples because:\\
- For a list like `[72, 88, 23, 92, 63, 100]`, the head `72` is not `100`, so it matches the recursive case and continues with the tail. The next head `88` is also not `100`, continuing to the next element. This process continues until it finds `100`, at which point it returns `100`.\\
- For an empty list `[]`, it directly matches the base case and returns `none`.\\
This definition does not cover any negative examples because there are no scenarios where `f/2` should return a value other than `100` or `none` based on the given positive examples.} \vspace{5pt}

\noindent
\textsl{\textbf{Prompt:} Now, please return only the code you have written in your last response. Answer only with the code, do not explain it.} \vspace{5pt}

\noindent
\textsl{\textbf{Answer:}}
\begin{small}
\begin{verbatim}
% Base case: if the list is empty, 
% return 'none'
f([], none).
% Recursive case: check if the 
% head of the list 
% is 100, and if so, return it; 
% otherwise, continue with the tail.
f([H|T], H) :- H =:= 100, !. 
% The cut (!) operator ensures that 
% only the first match is considered, 
% preventing backtracking into 
% non-matching tails.
f([H|T], Result) :- H =\= 100, f(T, Result).
\end{verbatim}
\end{small}\vspace{5pt}

\noindent
\textsl{\textbf{Prompt:} Your program is incorrect and has some errors. Here are errors and warnings from the Prolog interpreter:\\
Warning:~\dots/tmp.pl:5:\\
Warning:    Singleton variables: [T]\\
Additionally, your program still fails on the following examples:\\
Please think about how your program could have been wrong and provide a new Prolog program. Pay special attention to syntax!}
\vspace{5pt}

\noindent
\textsl{\textbf{Answer:}}
\textit{Apologies for the confusion earlier. Let's correct the Prolog program by ensuring that it handles both positive and negative cases properly, as well as addressing any syntax errors. Here is the revised version of the `f/2` predicate:}
\begin{small}
\begin{verbatim}
% Base case: if the list is empty, 
% return 'none'
f([], none).
% Recursive case: check if the head 
% of the list is 100, and if so,  
% return it; otherwise, continue 
% with the tail.
f([H|T], H) :- H =:= 100, !. 
% The cut (!) operator ensures that 
% only the first match is considered, 
% preventing backtracking into 
% non-matching tails.
f([H|T], Result) :- H =\= 100, f(T, Result).
\end{verbatim}\end{small}
\vspace{2pt}

\textit{This definition should now correctly cover both positive examples and not have any syntax errors or warnings:
\begin{enumerate}
    \item `pos(f([72,88,23,92,63,100],100))` is a positive example, which matches the recursive case where `H = 100`.
    \item `pos(f([],none))` is also a positive example, which matches the base case directly.
\end{enumerate}
If you still encounter any issues or warnings with this definition, please provide more details so I can assist you further.} \vspace{5pt}

\noindent
\textsl{\textbf{Prompt:} Now, please return only the code you have written in your last response. Answer only with the code, do not explain it.} 
\vspace{5pt}

\noindent
\textsl{\textbf{Answer:}}
\begin{small}
\begin{verbatim}
% Base case: if the list is empty, return 'none'
f([], none).
% Recursive case: check if the head of the list 
% is 100, and if so, return it; otherwise, 
% continue with the tail.
f([H|T], H) :- H =:= 100, !. 
% The cut (!) operator ensures that only the 
% first match is considered, preventing 
% backtracking into non-matching tails.
f([H|T], Result) :- H =\= 100, f(T, Result).
\end{verbatim}
\end{small}
\vspace{2pt}
This code revision step repeats at most two more times.

\subsection{Python}

\textsl{\textbf{Prompt:} Here are examples of the Python function f() with 1 parameters which takes a list as input and returns an element as output:\\
f([72,88,23,92,63,100]) $>>>$ 100\\
f([]) $>>>$ None\\
All examples are positive examples, meaning they show the correct behaviour of the function f().\\
Please explain which function f() would result in the provided examples and why. Please also provide Python code which defines the function.}
\vspace{5pt}

\noindent
\textsl{\textbf{Answer:}}
\textit{To create a Python function `f()` that behaves as described by your examples, we need to define what it means for `f()` to return an element from a list or None if the list is empty. A common interpretation of such functions in Python might be to find and return the maximum value in the list, or return None if the list is empty.\\
Here's how you can implement this:}
\begin{small}
\begin{lstlisting}[language=Python]
def f(lst):
    if len(lst) == 0:
        return None
    else:
        return max(lst)
\end{lstlisting}
\end{small}\vspace{2pt}
\textit{\textbf{Explanation:}
\begin{enumerate}
    \item \textbf{Function Definition}: The function f() takes one parameter, lst, which is expected to be a list of integers.
    \item \textbf{Condition Check}: Inside the function, we first check if the length of the list (len(lst)) is 0 using an if statement. If the list is empty, it returns `None`.
    \item \textbf{Return Maximum Value}: If the list is not empty, it uses Python's built-in max() function to find and return the largest number in the list.
\end{enumerate}
\textbf{Examples:}
\begin{itemize}
    \item For input [72, 88, 23, 92, 63, 100], the maximum value is 100, so it returns 100.
    \item For an empty list [], since there are no elements to find a maximum from, it returns None.
\end{itemize}
This function definition covers all positive examples provided (finding and returning the maximum element or returning None for an empty list).} \vspace{5pt}

\noindent
    \textsl{\textbf{Prompt:} Now, please return only the code you have written in your last response. Answer only with the code, do not explain it.}     

\vspace{5pt}

\noindent
\textsl{\textbf{Answer:}}
\begin{small}
\begin{lstlisting}[language=Python]
def f(lst):
    if len(lst) == 0:
        return None
    else:
        return max(lst)
\end{lstlisting}
\end{small}
\subsection{System Prompts}

We use two different prompts as the first instruction:
\begin{quote}
    \textsl{Your task is to learn a Prolog program from examples. Find the simplest program which covers all positive and no negative examples.}
\end{quote}

\begin{quote}
    \textsl{Your task is to learn a Python program from examples. Find the simplest program which covers all positive examples.}
\end{quote}

Then, we add the message history or current prompt. We use one prompt each for program generation, and another one for extracting the code from the examples.

% Note that due to the differences between the two programming languages, a Python function for \textsl{sorted} would return \textit{True} or \textit{False} depending on whether the input list is sorted. Since we explicitly state the return value which can be \textit{False} with correct function behavior, the negative examples in the Prolog case turn to positive examples with a different return value in the Python case.

% \begin{quote}
%     \textsl{Your task is to extract important information from natural language text for further processing.}
% \end{quote}

%% file: A-interface.tex
\section{Interface}
\label{sec:interface}
\begin{figure*}[t]
\includegraphics[width=0.85\textwidth]{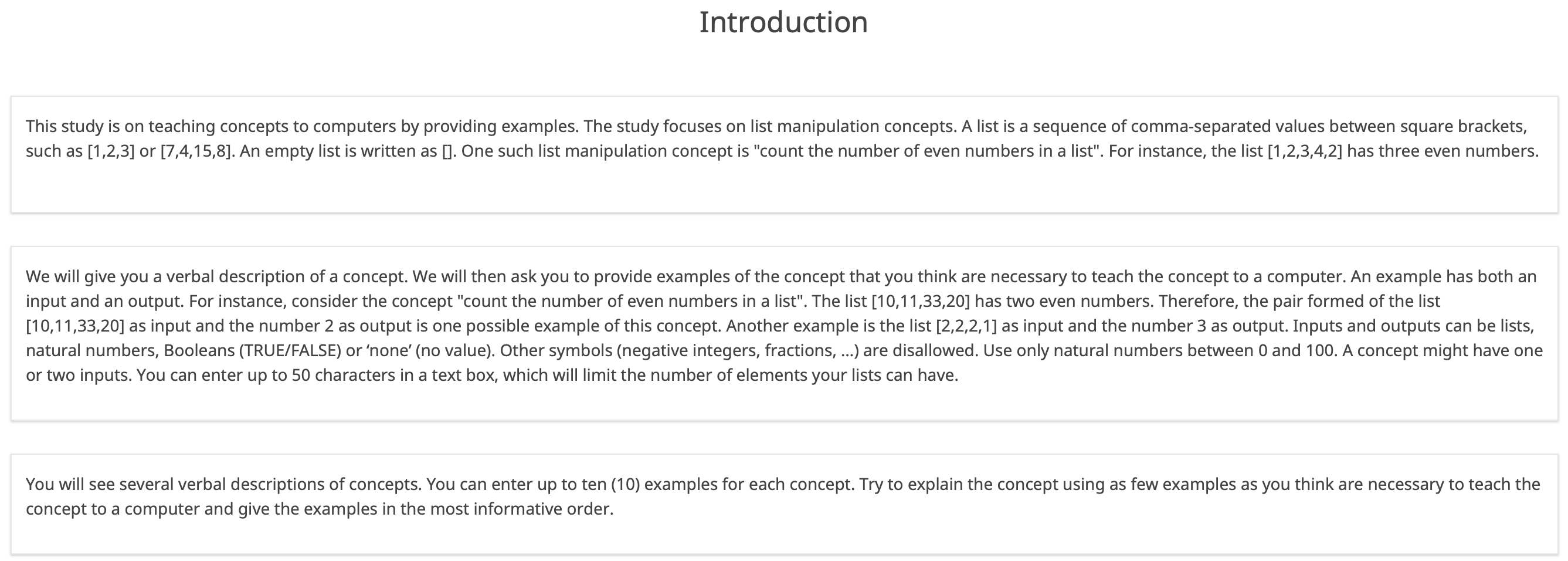}
\caption{Introductory text presented to participants.\label{fig:introduction}}
\end{figure*}

\begin{figure*}
\includegraphics[width=0.9\textwidth]{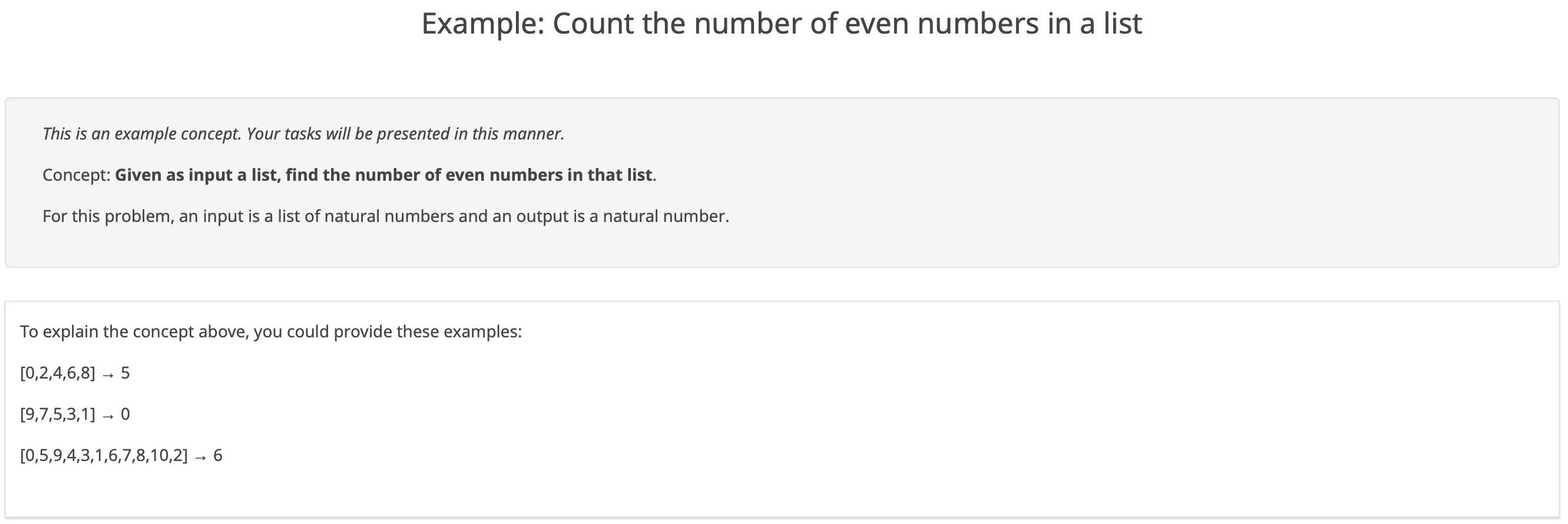}
\caption{An example of task: count the number of even numbers in a list.\label{fig:counteven}}
\end{figure*}

\begin{figure*}
\includegraphics[width=0.72\textwidth]{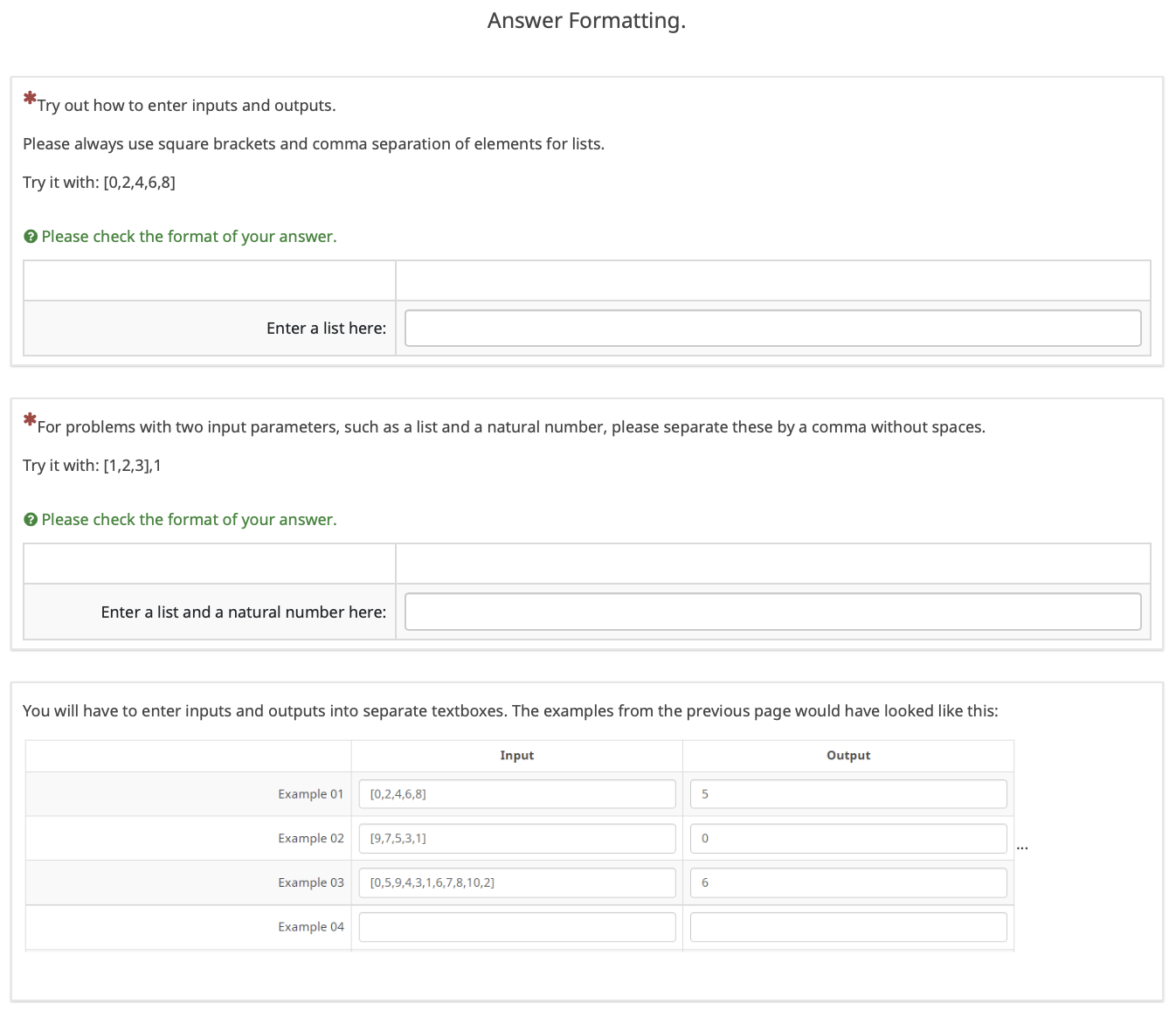}
\caption{Description of our answer formatting.
\label{fig:answerfomatting}}
\end{figure*}

\begin{figure*}
\includegraphics[width=0.72\textwidth]{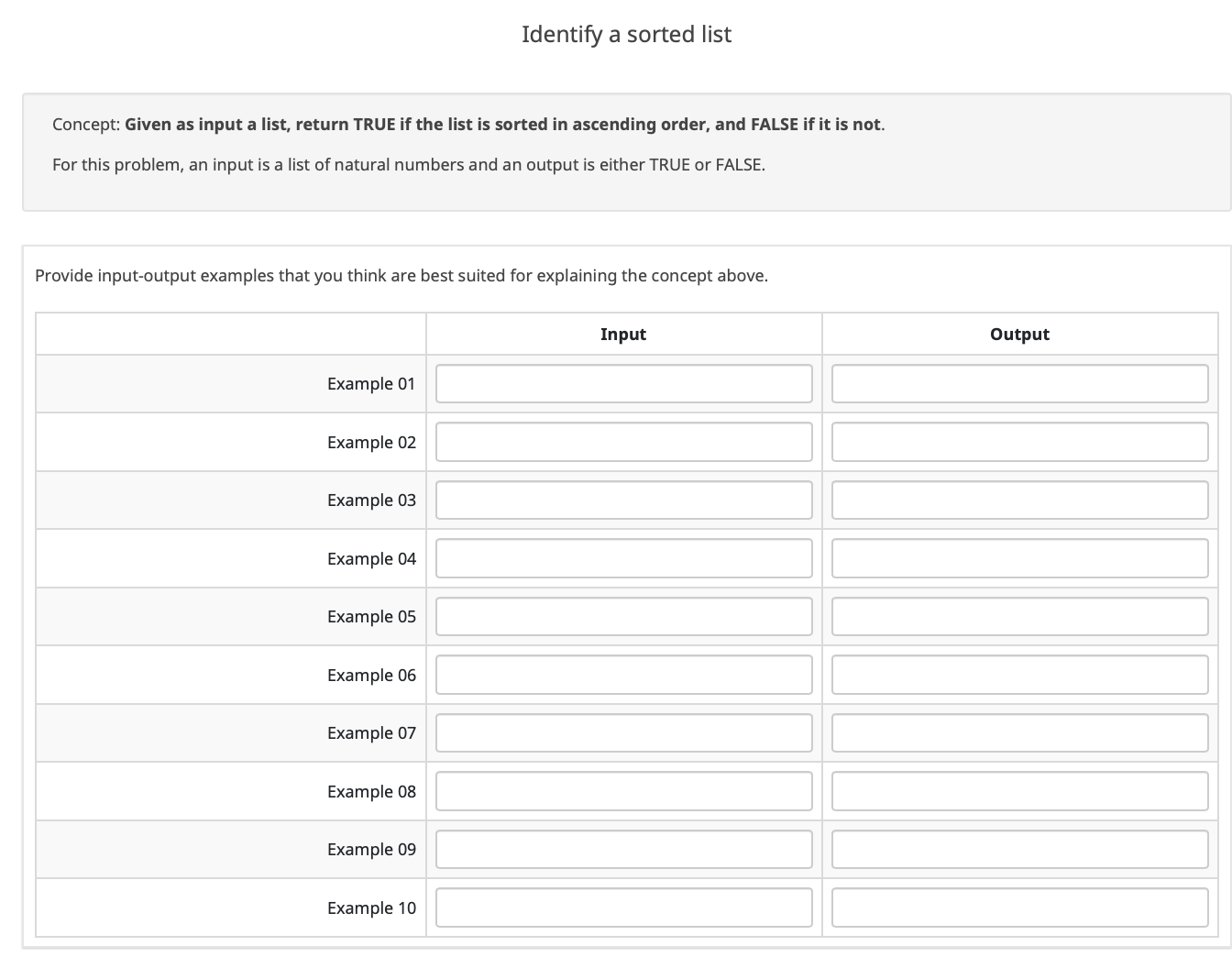}
\caption{Task \emph{sorted}.\label{fig:sorted}}
\end{figure*}

Figures \ref{fig:introduction} to \ref{fig:sorted} show our experimental interface. First, we showed participants an introductory text (Figure \ref{fig:introduction}). Then, we showed an example of task (Figure \ref{fig:counteven}). Next, we provided instructions on formatting answers (Figure \ref{fig:answerfomatting}). Finally, we presented six tasks, such as identifying a sorted list (Figure \ref{fig:sorted}).

% \begin{table}[ht]
% % \small
% \centering
% \footnotesize
% \begin{tabular}{@{}l|c@{}}
% % \toprule
% \textbf{Task} & Description \\\midrule
% \multirow{1}{*}{\emph{last}} & the last item of a list\\
% \multirow{1}{*}{\emph{last}} & the length of a list\\
% \multirow{1}{*}{\emph{append}} & appending an item to a list\\
% \multirow{1}{*}{\emph{maxlist}} & the maximum item of a list\\
% \multirow{1}{*}{\emph{dropk}} & drop first k items\\
% \multirow{1}{*}{\emph{sorted}} & identification of a sorted list
% \end{tabular}
% \caption{Verbal descriptions of the tasks shown to participants
% }
% \label{tab:verbal_description}
% \end{table}

%% file: B-expertexs.tex
\section{Expert examples}
\label{sec:expertexs}

Table \ref{tab:expert_exs} shows the examples provided by the expert.

\begin{table*}[ht]
% \small
\centering
\footnotesize
\begin{tabular}{@{}l|cc@{}}
% \toprule
\textbf{Task} & \textbf{Input} & \textbf{Output} \\\midrule
\multirow{2}{*}{\emph{last}} & $[72,88,23,92,63,100]$ & $100$ \\
 & $[]$ & $none$ \\\midrule
\multirow{1}{*}{\emph{length}} & $[72,88,23,2]$ & $4$ \\\midrule
\multirow{2}{*}{\emph{append}} & $[43, 99, 79],66$ & $[43, 99, 79, 66]$ \\
& $[70, 30, 72, 44, 35, 67, 58, 79, 96, 41], 56$ & $[70, 30, 72, 44, 35, 67, 58, 79, 96, 41, 56]$\\
\midrule
\multirow{3}{*}{\emph{maxlist}} & $[71,88,23,24,44,46,77,92,66,100,26,94,49,53]$ & $100$ \\
 & $[8,74,36,28,94,55,34,98,23,12,9,2]$  &  $98$ \\
  & $[]$ & $none$\\\midrule
\multirow{2}{*}{\emph{dropk}} & $[72,88,23,100,26,42,8,79,90],6$  &  $[8,79,90]$\\
 & $[23,45,87],0$  &  $[23,45,87]$ \\\midrule
 \multirow{4}{*}{\emph{sorted}} & $[1,2,4,9,13,26,39,42]$  &  $true$\\
 & $[1,2,4,9,26,25,39,42]$  &  $false$\\
& $[22,32,45,48,56,68,73,88]$  &  $true$\\
& $[]$  &  $true$\\
\end{tabular}
\caption{Examples provided by the expert}
\label{tab:expert_exs}
\end{table*}

%% file: C-results.tex
\section{Experimental results}\label{sec:appendixresults}

% \definecolor{mygreen}{rgb}{0.3,0.6,0.4}

%\ch{the curves below are not correct}

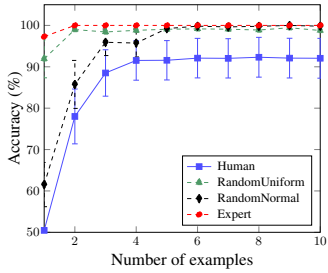
\begin{figure}
    \centering
\begin{tikzpicture}[scale=.53]
    \begin{axis}[
    xlabel=Number of examples,
    ylabel=Accuracy (\%),
    xmin=1,
    xmax=10,
    ymin=50,
    ylabel style={yshift=-3mm},
    label style={font=\Large},
    legend style={legend pos=south east,font=\normalsize,style={nodes={right}}}
    ]

\addplot+[blue!70,mark=square*,error bars/.cd,y fixed,y dir=both,y explicit] coordinates {
(1, 50.55714285714286) +- (0, 0.18655767917968585)
(2, 78.07857142857142) +- (0, 6.600858906578187)
(3, 88.64285714285715) +- (0, 5.541013061316756)
(4, 91.6392857142857) +- (0, 4.703322354415519)
(5, 91.63214285714287) +- (0, 4.7166283099068504)
(6, 92.16785714285714) +- (0, 4.726450027934498)
(7, 92.05714285714286) +- (0, 4.7383311186153865)
(8, 92.41071428571429) +- (0, 4.737233214882091)
(9, 92.1857142857143) +- (0, 4.736668476625405)
(10, 92.13571428571429) +- (0, 4.761984385781191)
};

\addplot+[black!30!green,mark=triangle*,mark options={scale=1.5},dashed,error bars/.cd,y fixed,y dir=both,y explicit] coordinates {
(1, 91.92857142857143) +- (0, 4.591480746710121)
(2, 98.99285714285713) +- (0, 0.3251312981973648)
(3, 98.41785714285714) +- (0, 0.384007466976721)
(4, 99.26071428571429) +- (0, 0.33142128876045696)
(5, 99.09285714285716) +- (0, 0.33975820195070855)
(6, 99.33214285714286) +- (0, 0.29527446455654155)
(7, 99.08571428571429) +- (0, 0.3421581702966988)
(8, 99.15000000000002) +- (0, 0.3186941073599163)
(9, 99.61071428571428) +- (0, 0.26737318917945707)
(10, 99.76785714285715) +- (0, 0.23214285714285698)
};

\addplot+[black!70,mark=diamond*,mark options={scale=1.5},dashed,error bars/.cd,y fixed,y dir=both,y explicit] coordinates {
(1, 61.807142857142864) +- (0, 5.316615199001553)
(2, 88.99285714285713) +- (0, 5.239699096780632)
(3, 95.93928571428572) +- (0, 3.2088398616099285)
(4, 99.14285714285714) +- (0, 0.324708539759549)
(5, 99.23928571428571) +- (0, 0.33603123920338895)
(6, 99.81785714285714) +- (0, 0.18214285714285694)
(7, 99.63928571428572) +- (0, 0.24507111389027939)
(8, 100.0) +- (0, 0.0)
(9, 100.0) +- (0, 0.0)
(10, 99.81428571428572) +- (0, 0.1857142857142859)
};

\addplot+[red!70,mark=otimes*,dashed,error bars/.cd,y fixed,y dir=both,y explicit] coordinates {
(1, 97.3) +- (0, 0)
(2, 100.0) +- (0, 0)
(3, 100.0) +- (0, 0)
(4, 100.0) +- (0, 0)
(5, 100.0) +- (0, 0)
(6, 100.0) +- (0, 0)
(7, 100.0) +- (0, 0)
(8, 100.0) +- (0, 0)
(9, 100.0) +- (0, 0)
(10, 100.0) +- (0, 0)
};

    \legend{Human,RandomUniform, RandomNormal,Expert}
    \end{axis}
\end{tikzpicture}
    \caption{Predictive accuracies for the NCS group for the task \emph{dropk}.}
\label{fig:dropkNCS}
\end{figure}

% Table \ref{tab:time} shows the detail of the time taken by the NCS and CS group to provide example sets for each of the tasks.
Figures \ref{fig:dropkNCS} to \ref{fig:lastNCS} show the detail of the results for each of the tasks for the NCS group.
Figures \ref{fig:dropkCS} to \ref{fig:lastCS} show the detail of the results for each of the tasks for the CS group.

% \begin{table}[ht]
% % \small
% \centering
% \footnotesize
% \begin{tabular}{@{}l|cc@{}}
% task & NCS & CS\\\midrule
% \emph{dropk} & 0min45s $\pm$ 0min36s & 0min45s $\pm$ 0min25s\\
% \emph{sorted} & 3min44s $\pm$ 1min30s & 4min27s $\pm$ 4min40s\\
% \emph{append} & 2min7s $\pm$ 0min52s & 2min24s $\pm$ 0min37s\\
% \emph{maxlist} & 2min21s $\pm$ 1min14s & 1min38s $\pm$ 0min36s\\
% \emph{length} & 1min32s $\pm$ 0min34s & 2min41s $\pm$ 1min57s\\
% \emph{last} & 1min44s $\pm$ 0min48s & 3min16s $\pm$ 4min2s\\
% \end{tabular}
% \caption{Time taken to provide example sets for the tasks
% }
% \label{tab:time}
% \end{table}

\begin{figure}
    \centering
\begin{tikzpicture}[scale=.53]
    \begin{axis}[
    xlabel=Number of examples,
    ylabel=Accuracy (\%),
    xmin=1,
    xmax=10,
    ymin=50,
    ylabel style={yshift=-3mm},
    label style={font=\Large},
    legend style={legend pos=south east,font=\normalsize,style={nodes={right}}}
    ]

\addplot+[blue!70,mark=square*,error bars/.cd,y fixed,y dir=both,y explicit] coordinates {
(1, 53.7) +- (0, 1.3805258593563718)
(2, 58.68928571428571) +- (0, 3.431440677847819)
(3, 62.37142857142857) +- (0, 2.414816412183873)
(4, 77.61785714285715) +- (0, 4.320992735264706)
(5, 78.09642857142858) +- (0, 4.235458199734174)
(6, 77.48571428571431) +- (0, 4.550040540126812)
(7, 77.37857142857143) +- (0, 4.890211444803002)
(8, 81.64285714285715) +- (0, 4.278979598575289)
(9, 83.27142857142856) +- (0, 4.3124459362297225)
(10, 84.85714285714285) +- (0, 4.539151591900663)
};

\addplot+[black!30!green,mark=triangle*,mark options={scale=1.5},dashed,error bars/.cd,y fixed,y dir=both,y explicit] coordinates {
(1, 69.51071428571429) +- (0, 5.1299839838969525)
(2, 65.47500000000001) +- (0, 4.2005943784026725)
(3, 69.28928571428571) +- (0, 3.2585988467852274)
(4, 72.14285714285712) +- (0, 2.495616250740008)
(5, 78.00357142857145) +- (0, 3.6820304072532055)
(6, 76.32142857142857) +- (0, 3.917138709209818)
(7, 84.08214285714286) +- (0, 4.621514448817394)
(8, 84.59642857142855) +- (0, 3.290714810477087)
(9, 82.92500000000001) +- (0, 3.206268542563609)
(10, 86.1285714285714) +- (0, 3.3268707630118968)
};

\addplot+[black!70,mark=diamond*,mark options={scale=1.5},dashed,error bars/.cd,y fixed,y dir=both,y explicit] coordinates {
(1, 54.01071428571428) +- (0, 0.7496002335939349)
(2, 60.0) +- (0, 3.809895300110254)
(3, 64.93928571428572) +- (0, 3.0143119172341653)
(4, 69.24285714285713) +- (0, 4.10697562965815)
(5, 81.70714285714286) +- (0, 4.6436026451347745)
(6, 82.22142857142856) +- (0, 4.500826280323592)
(7, 84.30714285714286) +- (0, 4.593126305923494)
(8, 94.39285714285714) +- (0, 2.7421459128247268)
(9, 94.58571428571429) +- (0, 3.491152194057459)
(10, 92.24285714285713) +- (0, 3.92275993732763)
};

\addplot+[red!70,mark=otimes*,dashed,error bars/.cd,y fixed,y dir=both,y explicit] coordinates {
(1, 50.0) +- (0, 0)
(2, 99.75) +- (0, 0)
(3, 99.7) +- (0, 0)
(4, 100.0) +- (0, 0)
(5, 100.0) +- (0, 0)
(6, 100.0) +- (0, 0)
(7, 100.0) +- (0, 0)
(8, 100.0) +- (0, 0)
(9, 100.0) +- (0, 0)
(10, 100.0) +- (0, 0)
};

    \legend{Human, RandomUniform, RandomNormal, Expert}
    \end{axis}
\end{tikzpicture}
    \caption{Predictive accuracies for the NCS group for the task \emph{sorted}.}
\label{fig:sortedNCS}
\end{figure}

\begin{figure}
    \centering
\begin{tikzpicture}[scale=.53]
    \begin{axis}[
    xlabel=Number of examples,
    ylabel=Accuracy (\%),
    xmin=1,
    xmax=10,
    ymin=50,
    ylabel style={yshift=-3mm},
    label style={font=\Large},
    legend style={legend pos=south east,font=\normalsize,style={nodes={right}}}
    ]

\addplot+[blue!70,mark=square*,error bars/.cd,y fixed,y dir=both,y explicit] coordinates {
(1, 77.025) +- (0, 4.668790316771361)
(2, 92.77142857142857) +- (0, 3.1708827864195697)
(3, 94.68571428571427) +- (0, 2.8227803132757727)
(4, 91.08571428571429) +- (0, 3.3190609545614236)
(5, 96.54642857142858) +- (0, 2.3482426459658745)
(6, 91.00714285714285) +- (0, 4.25146987233277)
(7, 94.63571428571429) +- (0, 2.84910545326607)
(8, 96.45) +- (0, 2.4121145709308056)
(9, 94.52499999999999) +- (0, 2.9101270222645415)
(10, 94.67142857142858) +- (0, 2.8306297317344358)
};

\addplot+[black!30!green,mark=triangle*,mark options={scale=1.5},dashed,error bars/.cd,y fixed,y dir=both,y explicit] coordinates {
(1, 100.0) +- (0, 0.0)
(2, 96.325) +- (0, 2.4966722357464652)
(3, 98.17857142857143) +- (0, 1.8214285714285718)
(4, 100.0) +- (0, 0.0)
(5, 100.0) +- (0, 0.0)
(6, 98.2) +- (0, 1.8000000000000003)
(7, 100.0) +- (0, 0.0)
(8, 100.0) +- (0, 0.0)
(9, 100.0) +- (0, 0.0)
(10, 100.0) +- (0, 0.0)
};

\addplot+[black!70,mark=diamond*,mark options={scale=1.5},dashed,error bars/.cd,y fixed,y dir=both,y explicit] coordinates {
(1, 99.3607142857143) +- (0, 0.6392857142857145)
(2, 100.0) +- (0, 0.0)
(3, 100.0) +- (0, 0.0)
(4, 100.0) +- (0, 0.0)
(5, 100.0) +- (0, 0.0)
(6, 100.0) +- (0, 0.0)
(7, 100.0) +- (0, 0.0)
(8, 100.0) +- (0, 0.0)
(9, 100.0) +- (0, 0.0)
(10, 100.0) +- (0, 0.0)
};

\addplot+[red!70,mark=otimes*,dashed,error bars/.cd,y fixed,y dir=both,y explicit] coordinates {
(1, 100.0) +- (0, 0)
(2, 100.0) +- (0, 0)
(3, 100.0) +- (0, 0)
(4, 100.0) +- (0, 0)
(5, 100.0) +- (0, 0)
(6, 100.0) +- (0, 0)
(7, 100.0) +- (0, 0)
(8, 100.0) +- (0, 0)
(9, 100.0) +- (0, 0)
(10, 100.0) +- (0, 0)
};

    \legend{Human, RandomUniform, RandomNormal, Expert}
    \end{axis}
\end{tikzpicture}
    \caption{Predictive accuracies for the NCS group for the task \emph{append}.}
\label{fig:appendNCS}
\end{figure}

\begin{figure}
    \centering
\begin{tikzpicture}[scale=.53]
    \begin{axis}[
    xlabel=Number of examples,
    ylabel=Accuracy (\%),
    xmin=1,
    xmax=10,
    ymin=50,
    ylabel style={yshift=-3mm},
    label style={font=\Large},
    legend style={legend pos=south east,font=\normalsize,style={nodes={right}}}
    ]

\addplot+[blue!70,mark=square*,error bars/.cd,y fixed,y dir=both,y explicit] coordinates {
(1, 51.12500000000001) +- (0, 0.39278345962355193)
(2, 75.28928571428571) +- (0, 6.71457161245418)
(3, 92.49285714285715) +- (0, 4.746394979995076)
(4, 92.5) +- (0, 4.750069403772891)
(5, 95.96428571428571) +- (0, 3.528227727176842)
(6, 95.98571428571428) +- (0, 3.5259611290867148)
(7, 95.96428571428571) +- (0, 3.5359434935141367)
(8, 96.01785714285714) +- (0, 3.5361461885168164)
(9, 95.9107142857143) +- (0, 3.5318495083687425)
(10, 96.04642857142856) +- (0, 3.492265646898677)
};

\addplot+[black!30!green,mark=triangle*,mark options={scale=1.5},dashed,error bars/.cd,y fixed,y dir=both,y explicit] coordinates {
(1, 87.99642857142857) +- (0, 5.171247283720607)
(2, 99.53571428571428) +- (0, 0.048727931445760334)
(3, 99.5642857142857) +- (0, 0.03758820082744391)
(4, 99.55357142857143) +- (0, 0.051479294278412996)
(5, 99.49642857142858) +- (0, 0.022522446504461047)
(6, 99.54642857142856) +- (0, 0.057527892354440405)
(7, 99.56428571428572) +- (0, 0.05846680551337437)
(8, 99.65) +- (0, 0.048038446141525935)
(9, 99.58928571428571) +- (0, 0.08020895983023732)
(10, 99.81071428571428) +- (0, 0.062277862696977444)
};

\addplot+[black!70,mark=diamond*,mark options={scale=1.5},dashed,error bars/.cd,y fixed,y dir=both,y explicit] coordinates {
(1, 85.97857142857143) +- (0, 5.823794772937989)
(2, 99.30000000000001) +- (0, 0.09721393650372065)
(3, 99.42857142857142) +- (0, 0.06413290752270306)
(4, 99.575) +- (0, 0.05415433784741826)
(5, 99.54642857142856) +- (0, 0.04553234742993568)
(6, 99.57142857142856) +- (0, 0.042164736358131304)
(7, 99.58571428571429) +- (0, 0.06471165363494859)
(8, 99.56785714285714) +- (0, 0.052833748906810606)
(9, 99.63214285714287) +- (0, 0.04933824243327893)
(10, 99.58928571428571) +- (0, 0.04893889760289971)
};

\addplot+[red!70,mark=otimes*,dashed,error bars/.cd,y fixed,y dir=both,y explicit] coordinates {
(1, 99.45) +- (0, 0)
(2, 99.55000000000001) +- (0, 0)
(3, 99.95) +- (0, 0)
(4, 100.0) +- (0, 0)
(5, 100.0) +- (0, 0)
(6, 100.0) +- (0, 0)
(7, 100.0) +- (0, 0)
(8, 100.0) +- (0, 0)
(9, 100.0) +- (0, 0)
(10, 100.0) +- (0, 0)
};
    \legend{Human, RandomUniform, RandomNormal, Expert}
    \end{axis}
\end{tikzpicture}
    \caption{Predictive accuracies for the NCS group for the task \emph{maxlist}.}
\label{fig:maxNCS}
\end{figure}

\begin{figure}
    \centering
\begin{tikzpicture}[scale=.53]
    \begin{axis}[
    xlabel=Number of examples,
    ylabel=Accuracy (\%),
    xmin=1,
    xmax=10,
    ymin=50,
    ylabel style={yshift=-3mm},
    label style={font=\Large},
    legend style={legend pos=south east,font=\normalsize,style={nodes={right}}}
    ]

\addplot+[blue!70,mark=square*,error bars/.cd,y fixed,y dir=both,y explicit] coordinates {
(1, 64.55357142857142) +- (0, 6.218002630358563)
(2, 92.85714285714286) +- (0, 4.852615860619696)
(3, 92.93571428571428) +- (0, 4.799248534845333)
(4, 96.43928571428572) +- (0, 3.5607142857142864)
(5, 96.42857142857143) +- (0, 3.571428571428571)
(6, 100.0) +- (0, 0.0)
(7, 100.0) +- (0, 0.0)
(8, 100.0) +- (0, 0.0)
(9, 100.0) +- (0, 0.0)
(10, 100.0) +- (0, 0.0)
};

\addplot+[black!30!green,mark=triangle*,mark options={scale=1.5},dashed,error bars/.cd,y fixed,y dir=both,y explicit] coordinates {
(1, 92.86428571428571) +- (0, 4.84776607828436)
(2, 100.0) +- (0, 0.0)
(3, 100.0) +- (0, 0.0)
(4, 100.0) +- (0, 0.0)
(5, 100.0) +- (0, 0.0)
(6, 100.0) +- (0, 0.0)
(7, 100.0) +- (0, 0.0)
(8, 100.0) +- (0, 0.0)
(9, 100.0) +- (0, 0.0)
(10, 100.0) +- (0, 0.0)
};

\addplot+[black!70,mark=diamond*,mark options={scale=1.5},dashed,error bars/.cd,y fixed,y dir=both,y explicit] coordinates {
(1, 94.77857142857144) +- (0, 3.7465138164806038)
(2, 100.0) +- (0, 0.0)
(3, 100.0) +- (0, 0.0)
(4, 100.0) +- (0, 0.0)
(5, 100.0) +- (0, 0.0)
(6, 100.0) +- (0, 0.0)
(7, 100.0) +- (0, 0.0)
(8, 100.0) +- (0, 0.0)
(9, 100.0) +- (0, 0.0)
(10, 100.0) +- (0, 0.0)
};

\addplot+[red!70,mark=otimes*,dashed,error bars/.cd,y fixed,y dir=both,y explicit] coordinates {
(1, 100.0) +- (0, 0)
(2, 100.0) +- (0, 0)
(3, 100.0) +- (0, 0)
(4, 100.0) +- (0, 0)
(5, 100.0) +- (0, 0)
(6, 100.0) +- (0, 0)
(7, 100.0) +- (0, 0)
(8, 100.0) +- (0, 0)
(9, 100.0) +- (0, 0)
(10, 100.0) +- (0, 0)
};

    \legend{Human, RandomUniform, RandomNormal, Expert}
    \end{axis}
\end{tikzpicture}
    \caption{Predictive accuracies for the NCS group for the task \emph{length}.}
\label{fig:lengthNCS}
\end{figure}

\begin{figure}
    \centering
\begin{tikzpicture}[scale=.53]
    \begin{axis}[
    xlabel=Number of examples,
    ylabel=Accuracy (\%),
    xmin=1,
    xmax=10,
    ymin=50,
    ylabel style={yshift=-3mm},
    label style={font=\Large},
    legend style={legend pos=south east,font=\normalsize,style={nodes={right}}}
    ]

\addplot+[blue!70,mark=square*,error bars/.cd,y fixed,y dir=both,y explicit] coordinates {
(1, 64.21071428571429) +- (0, 6.170726217283152)
(2, 67.72142857142856) +- (0, 6.581846734497904)
(3, 81.87499999999999) +- (0, 6.581128844516865)
(4, 81.82142857142857) +- (0, 6.578339653002639)
(5, 81.85357142857143) +- (0, 6.584953011350608)
(6, 88.94642857142857) +- (0, 5.624269096068587)
(7, 88.95) +- (0, 5.641649961289188)
(8, 88.86428571428571) +- (0, 5.629283298543345)
(9, 88.92142857142858) +- (0, 5.637513466482575)
(10, 88.86785714285715) +- (0, 5.629747342580065)
};

\addplot+[black!30!green,mark=triangle*,mark options={scale=1.5},dashed,error bars/.cd,y fixed,y dir=both,y explicit] coordinates {
(1, 95.84285714285716) +- (0, 3.5274117193759196)
(2, 99.50714285714287) +- (0, 0.03737879417954941)
(3, 99.57142857142857) +- (0, 0.04738035414793448)
(4, 99.57142857142857) +- (0, 0.0523393107872962)
(5, 99.52857142857144) +- (0, 0.035438495596916704)
(6, 99.53928571428571) +- (0, 0.042960041307435456)
(7, 99.56785714285714) +- (0, 0.0660101438377612)
(8, 99.58214285714287) +- (0, 0.056599407555519385)
(9, 99.56428571428572) +- (0, 0.0404061017820884)
(10, 99.51785714285715) +- (0, 0.06324710453719656)
};

\addplot+[black!70,mark=diamond*,mark options={scale=1.5},dashed,error bars/.cd,y fixed,y dir=both,y explicit] coordinates {
(1, 85.71428571428571) +- (0, 6.103069459615007)
(2, 92.48571428571431) +- (0, 4.774058300132208)
(3, 99.53571428571428) +- (0, 0.03902259539464781)
(4, 99.58928571428571) +- (0, 0.045747326658388786)
(5, 99.57142857142857) +- (0, 0.033852575386262206)
(6, 99.57857142857144) +- (0, 0.05806265224356617)
(7, 99.63571428571427) +- (0, 0.049288501274919794)
(8, 99.61785714285713) +- (0, 0.06496134583057327)
(9, 99.60714285714286) +- (0, 0.05518628364610896)
(10, 99.57857142857142) +- (0, 0.05734853095886324)
};

\addplot+[red!70,mark=otimes*,dashed,error bars/.cd,y fixed,y dir=both,y explicit] coordinates {
(1,99.65) +- (0,0)
(2,100.0) +- (0,0)
(3,100.0) +- (0,0)
(4,100.0) +- (0,0)
(5,100.0) +- (0,0)
(6,100.0) +- (0,0)
(7,100.0) +- (0,0)
(8,100.0) +- (0,0)
(9,100.0) +- (0,0)
(10,100.0) +- (0,0)
};

    \legend{Human, RandomUniform, RandomNormal, Expert}
    \end{axis}
\end{tikzpicture}
    \caption{Predictive accuracies for the NCS group for the task \emph{last}.}
\label{fig:lastNCS}
\end{figure}

\newpage

\begin{figure}
    \centering
\begin{tikzpicture}[scale=.53]
    \begin{axis}[
    xlabel=Number of examples,
    ylabel=Accuracy (\%),
    xmin=1,
    xmax=10,
    ymin=50,
    ylabel style={yshift=-3mm},
    label style={font=\Large},
    legend style={legend pos=south east,font=\normalsize,style={nodes={right}}}
    ]

\addplot+[blue!70,mark=square*,error bars/.cd,y fixed,y dir=both,y explicit] coordinates {
(1, 52.23800000000001) +- (0, 1.8796038589731259)
(2, 56.152) +- (0, 3.2478909567081637)
(3, 66.014) +- (0, 4.712536931773939)
(4, 71.82799999999999) +- (0, 4.939343816068418)
(5, 69.944) +- (0, 4.923246997663229)
(6, 72.03600000000002) +- (0, 4.977904110500054)
(7, 70.04400000000001) +- (0, 4.947024627659202)
(8, 69.776) +- (0, 4.910752759676124)
(9, 69.914) +- (0, 4.92501566833379)
(10, 66.176) +- (0, 4.687860315040683)
};

\addplot+[black!30!green,mark=triangle*,mark options={scale=1.5},dashed,error bars/.cd,y fixed,y dir=both,y explicit] coordinates {
(1, 64.3) +- (0, 4.274020355590272)
(2, 84.812) +- (0, 4.341863348072268)
(3, 90.558) +- (0, 3.5926680893174643)
(4, 89.5) +- (0, 3.96767984259483)
(5, 84.006) +- (0, 4.3154174382864365)
(6, 78.19800000000001) +- (0, 4.819333875132538)
(7, 83.892) +- (0, 4.494791133449176)
(8, 93.268) +- (0, 3.216837059390274)
(9, 92.10799999999999) +- (0, 3.338647031358661)
(10, 88.334) +- (0, 4.00883578278449)
};

\addplot+[black!70,mark=diamond*,mark options={scale=1.5},dashed,error bars/.cd,y fixed,y dir=both,y explicit] coordinates {
(1, 55.016) +- (0, 2.575122521356994)
(2, 74.04400000000001) +- (0, 4.761838160486628)
(3, 83.65799999999999) +- (0, 4.529376042385234)
(4, 78.196) +- (0, 4.836357892739811)
(5, 87.72599999999998) +- (0, 4.207779620338181)
(6, 87.838) +- (0, 4.219599388567592)
(7, 85.75800000000001) +- (0, 4.413692407346333)
(8, 86.008) +- (0, 4.4072824960512795)
(9, 97.512) +- (0, 1.926761704691752)
(10, 90.018) +- (0, 3.954489937611002)
};

\addplot+[red!70,mark=otimes*,dashed,error bars/.cd,y fixed,y dir=both,y explicit] coordinates {
(1, 97.3) +- (0, 0)
(2, 100.0) +- (0, 0)
(3, 100.0) +- (0, 0)
(4, 100.0) +- (0, 0)
(5, 100.0) +- (0, 0)
(6, 100.0) +- (0, 0)
(7, 100.0) +- (0, 0)
(8, 100.0) +- (0, 0)
(9, 100.0) +- (0, 0)
(10, 100.0) +- (0, 0)

};

    \legend{Human, RandomUniform, RandomNormal, Expert}
    \end{axis}
\end{tikzpicture}
    \caption{Predictive accuracies for the CS group for the task \emph{dropk}.}
\label{fig:dropkCS}
\end{figure}

\begin{figure}
    \centering
\begin{tikzpicture}[scale=.53]
    \begin{axis}[
    xlabel=Number of examples,
    ylabel=Accuracy (\%),
    xmin=1,
    xmax=10,
    ymin=50,
    ylabel style={yshift=-3mm},
    label style={font=\Large},
    legend style={legend pos=south east,font=\normalsize,style={nodes={right}}}
    ]

\addplot+[blue!70,mark=square*,error bars/.cd,y fixed,y dir=both,y explicit] coordinates {
(1, 50.57000000000001) +- (0, 0.3190741815523989)
(2, 55.59) +- (0, 2.0968388906478563)
(3, 56.517999999999994) +- (0, 1.7564497336008986)
(4, 60.61) +- (0, 2.529733187512074)
(5, 58.897999999999996) +- (0, 2.294859254943536)
(6, 67.098) +- (0, 3.136305522957439)
(7, 68.46000000000001) +- (0, 3.285462930344317)
(8, 70.51) +- (0, 3.4185474498193917)
(9, 73.91400000000002) +- (0, 3.9073067450611036)
(10, 74.20400000000001) +- (0, 3.86083855312979)
};

\addplot+[black!30!green,mark=triangle*,mark options={scale=1.5},dashed,error bars/.cd,y fixed,y dir=both,y explicit] coordinates {
(1, 66.64800000000001) +- (0, 4.161640782191562)
(2, 64.372) +- (0, 2.5360666395029923)
(3, 68.25399999999999) +- (0, 2.0568339748263593)
(4, 67.656) +- (0, 2.391414016852791)
(5, 74.656) +- (0, 2.810304431907689)
(6, 74.618) +- (0, 2.751190469596753)
(7, 78.09) +- (0, 2.97474508711811)
(8, 84.20400000000001) +- (0, 2.8290736410823007)
(9, 85.844) +- (0, 3.1597828511887758)
(10, 85.45) +- (0, 2.328857015791223)
};

\addplot+[black!70,mark=diamond*,mark options={scale=1.5},dashed,error bars/.cd,y fixed,y dir=both,y explicit] coordinates {
(1, 52.736000000000004) +- (0, 0.48823252656905197)
(2, 58.224) +- (0, 2.0676136002648073)
(3, 61.248000000000005) +- (0, 2.135695593165531)
(4, 59.97) +- (0, 2.3180253809366858)
(5, 71.242) +- (0, 2.3938631261345473)
(6, 78.108) +- (0, 3.8785636086915822)
(7, 89.44999999999999) +- (0, 3.202608311985716)
(8, 88.00200000000001) +- (0, 3.6138027985304717)
(9, 88.596) +- (0, 3.4096174272196587)
(10, 95.05000000000001) +- (0, 2.153669349428242)
};

\addplot+[red!70,mark=otimes*,dashed,error bars/.cd,y fixed,y dir=both,y explicit] coordinates {
(1, 50.0) +- (0, 0)
(2, 99.75) +- (0, 0)
(3, 99.7) +- (0, 0)
(4, 100.0) +- (0, 0)
(5, 100.0) +- (0, 0)
(6, 100.0) +- (0, 0)
(7, 100.0) +- (0, 0)
(8, 100.0) +- (0, 0)
(9, 100.0) +- (0, 0)
(10, 100.0) +- (0, 0)
};

    \legend{Human, RandomUniform, RandomNormal, Expert}
    \end{axis}
\end{tikzpicture}
    \caption{Predictive accuracies for the CS group for the task \emph{sorted}.}
\label{fig:sortedCS}
\end{figure}

\begin{figure}
    \centering
\begin{tikzpicture}[scale=.53]
    \begin{axis}[
    xlabel=Number of examples,
    ylabel=Accuracy (\%),
    xmin=1,
    xmax=10,
    ymin=50,
    ylabel style={yshift=-3mm},
    label style={font=\Large},
    legend style={legend pos=south east,font=\normalsize,style={nodes={right}}}
    ]
\addplot+[blue!70,mark=square*,error bars/.cd,y fixed,y dir=both,y explicit] coordinates {
(1, 80.162) +- (0, 4.959633118151651)
(2, 92.066) +- (0, 3.7109553127642663)
(3, 96.0) +- (0, 2.7688746209726918)
(4, 96.0) +- (0, 2.7688746209726918)
(5, 96.0) +- (0, 2.7688746209726918)
(6, 96.0) +- (0, 2.7688746209726918)
(7, 94.0) +- (0, 3.3166247903554007)
(8, 94.986) +- (0, 2.8908238733389946)
(9, 94.964) +- (0, 2.8973486730687648)
(10, 94.96000000000001) +- (0, 2.8985513623187704)
};

\addplot+[black!30!green,mark=triangle*,mark options={scale=1.5},dashed,error bars/.cd,y fixed,y dir=both,y explicit] coordinates {
(1, 98.94) +- (0, 1.0600000000000003)
(2, 100.0) +- (0, 0.0)
(3, 100.0) +- (0, 0.0)
(4, 100.0) +- (0, 0.0)
(5, 100.0) +- (0, 0.0)
(6, 100.0) +- (0, 0.0)
(7, 100.0) +- (0, 0.0)
(8, 100.0) +- (0, 0.0)
(9, 100.0) +- (0, 0.0)
(10, 100.0) +- (0, 0.0)

};

\addplot+[black!70,mark=diamond*,mark options={scale=1.5},dashed,error bars/.cd,y fixed,y dir=both,y explicit] coordinates {
(1, 100.0) +- (0, 0.0)
(2, 100.0) +- (0, 0.0)
(3, 100.0) +- (0, 0.0)
(4, 100.0) +- (0, 0.0)
(5, 100.0) +- (0, 0.0)
(6, 100.0) +- (0, 0.0)
(7, 100.0) +- (0, 0.0)
(8, 100.0) +- (0, 0.0)
(9, 100.0) +- (0, 0.0)
(10, 100.0) +- (0, 0.0)
};

\addplot+[red!70,mark=otimes*,dashed,error bars/.cd,y fixed,y dir=both,y explicit] coordinates {
(1, 100.0) +- (0, 0)
(2, 100.0) +- (0, 0)
(3, 100.0) +- (0, 0)
(4, 100.0) +- (0, 0)
(5, 100.0) +- (0, 0)
(6, 100.0) +- (0, 0)
(7, 100.0) +- (0, 0)
(8, 100.0) +- (0, 0)
(9, 100.0) +- (0, 0)
(10, 100.0) +- (0, 0)
};

    \legend{Human, RandomUniform, RandomNormal, Expert}
    \end{axis}
\end{tikzpicture}
    \caption{Predictive accuracies for the CS group for the task \emph{append}.}
\label{fig:appendCS}
\end{figure}

\begin{figure}
    \centering
\begin{tikzpicture}[scale=.53]
    \begin{axis}[
    xlabel=Number of examples,
    ylabel=Accuracy (\%),
    xmin=1,
    xmax=10,
    ymin=50,
    ylabel style={yshift=-3mm},
    label style={font=\Large},
    legend style={legend pos=south east,font=\normalsize,style={nodes={right}}}
    ]

\addplot+[blue!70,mark=square*,error bars/.cd,y fixed,y dir=both,y explicit] coordinates {
(1, 51.284) +- (0, 0.3373885001004036)
(2, 56.928) +- (0, 3.2236820045821313)
(3, 60.854) +- (0, 3.978183128012083)
(4, 72.358) +- (0, 4.920236003824749)
(5, 76.178) +- (0, 4.963122404293491)
(6, 76.032) +- (0, 5.014173145527918)
(7, 77.72399999999999) +- (0, 4.932053426312412)
(8, 77.74600000000001) +- (0, 4.925699544227196)
(9, 78.78) +- (0, 4.8617126611925565)
(10, 78.798) +- (0, 4.849956597743944)
};

\addplot+[black!30!green,mark=triangle*,mark options={scale=1.5},dashed,error bars/.cd,y fixed,y dir=both,y explicit] coordinates {
(1, 83.064) +- (0, 4.4521217039369745)
(2, 99.444) +- (0, 0.062002688113769217)
(3, 99.55799999999999) +- (0, 0.033625387234449286)
(4, 99.528) +- (0, 0.03150132272355137)
(5, 99.522) +- (0, 0.035949038002891245)
(6, 99.58) +- (0, 0.04183300132670361)
(7, 99.61200000000001) +- (0, 0.045118362854459325)
(8, 99.572) +- (0, 0.04397726685459194)
(9, 99.654) +- (0, 0.050901866370497574)
(10, 99.68) +- (0, 0.059371710435189344)

};

\addplot+[black!70,mark=diamond*,mark options={scale=1.5},dashed,error bars/.cd,y fixed,y dir=both,y explicit] coordinates {
(1, 86.01799999999999) +- (0, 4.321769390731841)
(2, 97.32400000000001) +- (0, 1.9107745375457916)
(3, 99.376) +- (0, 0.060008332754709875)
(4, 99.47800000000001) +- (0, 0.047878317987720874)
(5, 99.58) +- (0, 0.03582364210034097)
(6, 99.54199999999999) +- (0, 0.048276978640617724)
(7, 99.558) +- (0, 0.05482396069846328)
(8, 99.606) +- (0, 0.04206344414492636)
(9, 99.55799999999999) +- (0, 0.04336665385600624)
(10, 99.602) +- (0, 0.04738846554454634)
};

\addplot+[red!70,mark=otimes*,dashed,error bars/.cd,y fixed,y dir=both,y explicit] coordinates {
(1, 99.45) +- (0, 0)
(2, 99.55000000000001) +- (0, 0)
(3, 99.95) +- (0, 0)
(4, 100.0) +- (0, 0)
(5, 100.0) +- (0, 0)
(6, 100.0) +- (0, 0)
(7, 100.0) +- (0, 0)
(8, 100.0) +- (0, 0)
(9, 100.0) +- (0, 0)
(10, 100.0) +- (0, 0)
};

    \legend{Human, RandomUniform, RandomNormal, Expert}
    \end{axis}
\end{tikzpicture}
    \caption{Predictive accuracies for the CS group for the task \emph{maxlist}.}
    \label{fig:maxlistCS}
\end{figure}

\begin{figure}
    \centering
\begin{tikzpicture}[scale=.53]
    \begin{axis}[
    xlabel=Number of examples,
    ylabel=Accuracy (\%),
    xmin=1,
    xmax=10,
    ymin=50,
    ylabel style={yshift=-3mm},
    label style={font=\Large},
    legend style={legend pos=south east,font=\normalsize,style={nodes={right}}}
    ]

\addplot+[blue!70,mark=square*,error bars/.cd,y fixed,y dir=both,y explicit] coordinates {
(1, 56.38999999999999) +- (0, 3.2878082871927514)
(2, 70.40399999999998) +- (0, 4.933111695471734)
(3, 92.07) +- (0, 3.709002561336403)
(4, 92.03399999999999) +- (0, 3.7258049689518997)
(5, 94.074) +- (0, 3.2760635931963633)
(6, 96.06) +- (0, 2.7273476126082645)
(7, 96.03) +- (0, 2.7481721197916262)
(8, 94.13) +- (0, 3.244824083572688)
(9, 94.074) +- (0, 3.2757328950938596)
(10, 94.192) +- (0, 3.2119732771823206)
};

\addplot+[black!30!green,mark=triangle*,mark options={scale=1.5},dashed,error bars/.cd,y fixed,y dir=both,y explicit] coordinates {
(1, 88.062) +- (0, 4.085017625420973)
(2, 98.0) +- (0, 2.0)
(3, 100.0) +- (0, 0.0)
(4, 100.0) +- (0, 0.0)
(5, 100.0) +- (0, 0.0)
(6, 100.0) +- (0, 0.0)
(7, 100.0) +- (0, 0.0)
(8, 100.0) +- (0, 0.0)
(9, 100.0) +- (0, 0.0)
(10, 100.0) +- (0, 0.0)
};

\addplot+[black!70,mark=diamond*,mark options={scale=1.5},dashed,error bars/.cd,y fixed,y dir=both,y explicit] coordinates {
(1, 87.258) +- (0, 4.274636124864899)
(2, 100.0) +- (0, 0.0)
(3, 100.0) +- (0, 0.0)
(4, 100.0) +- (0, 0.0)
(5, 100.0) +- (0, 0.0)
(6, 100.0) +- (0, 0.0)
(7, 100.0) +- (0, 0.0)
(8, 100.0) +- (0, 0.0)
(9, 100.0) +- (0, 0.0)
(10, 100.0) +- (0, 0.0)
};

\addplot+[red!70,mark=otimes*,dashed,error bars/.cd,y fixed,y dir=both,y explicit] coordinates {
(1, 100.0) +- (0, 0)
(2, 100.0) +- (0, 0)
(3, 100.0) +- (0, 0)
(4, 100.0) +- (0, 0)
(5, 100.0) +- (0, 0)
(6, 100.0) +- (0, 0)
(7, 100.0) +- (0, 0)
(8, 100.0) +- (0, 0)
(9, 100.0) +- (0, 0)
(10, 100.0) +- (0, 0)
};

    \legend{Human, RandomUniform, RandomNormal, Expert}
    \end{axis}
\end{tikzpicture}
    \caption{Predictive accuracies for the CS group for the task \emph{length}.}
\label{fig:lengthCS}
\end{figure}

\begin{figure}
    \centering
\begin{tikzpicture}[scale=.53]
    \begin{axis}[
    xlabel=Number of examples,
    ylabel=Accuracy (\%),
    xmin=1,
    xmax=10,
    ymin=50,
    ylabel style={yshift=-3mm},
    label style={font=\Large},
    legend style={legend pos=south east,font=\normalsize,style={nodes={right}}}
    ]

\addplot+[blue!70,mark=square*,error bars/.cd,y fixed,y dir=both,y explicit] coordinates {
(1, 58.282000000000004) +- (0, 3.6756846074348295)
(2, 70.11200000000001) +- (0, 4.893564890070769)
(3, 83.77600000000001) +- (0, 4.684794659320727)
(4, 87.726) +- (0, 4.324829592018626)
(5, 90.94) +- (0, 3.7073699123053085)
(6, 92.92600000000002) +- (0, 3.314121452210224)
(7, 92.97399999999999) +- (0, 3.3150969317150696)
(8, 92.906) +- (0, 3.3175924001199024)
(9, 92.968) +- (0, 3.3141274366968245)
(10, 92.96600000000001) +- (0, 3.3135362781576223)
};

\addplot+[black!30!green,mark=triangle*,mark options={scale=1.5},dashed,error bars/.cd,y fixed,y dir=both,y explicit] coordinates {
(1, 96.33999999999999) +- (0, 2.1853222493109192)
(2, 99.546) +- (0, 0.033901819026909676)
(3, 99.552) +- (0, 0.03668332954717536)
(4, 99.56) +- (0, 0.04051748593714405)
(5, 99.544) +- (0, 0.022045407685048483)
(6, 99.554) +- (0, 0.04855580981372523)
(7, 99.56) +- (0, 0.03989569734528605)
(8, 99.58000000000001) +- (0, 0.04462809279665284)
(9, 99.518) +- (0, 0.06047589051294195)
(10, 99.614) +- (0, 0.03323151115031215)

};

\addplot+[black!70,mark=diamond*,mark options={scale=1.5},dashed,error bars/.cd,y fixed,y dir=both,y explicit] coordinates {
(1, 70.24399999999999) +- (0, 4.869812898527691)
(2, 81.85600000000001) +- (0, 4.770339191294473)
(3, 87.69000000000001) +- (0, 4.285362683989925)
(4, 93.66799999999999) +- (0, 3.2480967247502552)
(5, 91.674) +- (0, 3.675131652970634)
(6, 91.72999999999999) +- (0, 3.698942416421213)
(7, 93.67999999999999) +- (0, 3.252691193458119)
(8, 91.72600000000001) +- (0, 3.687487988681003)
(9, 93.684) +- (0, 3.2654322327475525)
(10, 97.61) +- (0, 1.9735817692712911)
};

\addplot+[red!70,mark=otimes*,dashed,error bars/.cd,y fixed,y dir=both,y explicit] coordinates {
(1, 99.65) +- (0, 0)
(2, 100.0) +- (0, 0)
(3, 100.0) +- (0, 0)
(4, 100.0) +- (0, 0)
(5, 100.0) +- (0, 0)
(6, 100.0) +- (0, 0)
(7, 100.0) +- (0, 0)
(8, 100.0) +- (0, 0)
(9, 100.0) +- (0, 0)
(10, 100.0) +- (0, 0)
};

    \legend{Human, RandomUniform, RandomNormal, Expert}
    \end{axis}
\end{tikzpicture}
    \caption{Predictive accuracies for the CS group for the task \emph{last}.}
\label{fig:lastCS}
\end{figure}